\documentclass[twocolumn,unsortedaddress,showpacs,preprintnumbers,amsmath,amssymb]{revtex4-1}%

\usepackage{graphicx}
\usepackage{dcolumn}
\usepackage{bm}
\usepackage{subfigure}

\begin{document}

\preprint{APS/2012}

%\title{A mesoscopic model for microscale flows with interfacial effects: \\
%Slip, films, and contact angle hysteresis.}

\title{A mesoscopic model for microscale hydrodynamics and interfacial phenomena:\\ Slip, films, and contact angle hysteresis.}

\author{Carlos E. Colosqui}
\email{colosqui@princeton.edu} 
\affiliation{Department of Chemical and Biological Engineering,
Princeton University, Princeton, NJ 08544, USA}

\author{Michail E. Kavousanakis}
\affiliation{School of Chemical Engineering, National Technical
University of Athens, 15780, Greece}

\author{Athanasios G. Papathanasiou}
\affiliation{School of Chemical Engineering, National Technical University of Athens, 15780, Greece}

\author{Ioannis G. Kevrekidis}
\affiliation{Department of Chemical and Biological Engineering\\ and Program in Applied and Computational Mathematics,
Princeton University, Princeton, NJ 08544, USA}

\date{January 2012}

\begin{abstract}

We present a model based on the lattice Boltzmann equation that is suitable for the simulation of dynamic wetting.
The model is capable of exhibiting fundamental interfacial phenomena such as weak adsorption of fluid on the solid substrate and the presence of a thin surface film within which a disjoining pressure acts. 
Dynamics in this surface film, tightly coupled with hydrodynamics in the fluid bulk, determine macroscopic properties of primary interest: the hydrodynamic slip; the equilibrium contact angle; and the static and dynamic hysteresis of the contact angles.
The pseudo-potentials employed for fluid-solid interactions are composed of a repulsive core and an attractive tail that can be independently adjusted.
This enables effective modification of the functional form of the disjoining pressure so that one can vary the static and dynamic hysteresis on surfaces that exhibit the same equilibrium contact angle.
The modeled solid-fluid interface is diffuse, represented by a wall probability function which ultimately controls the momentum exchange between solid and fluid phases.
This approach allows us to effectively vary the slip length for a given wettability (i.e. the static contact angle) of the solid substrate.
\end{abstract}

\pacs{47.85.-g;47.55.-t;68.08.Bc}

%\keywords{}
\maketitle

%*****************************************************************************
% Section 1: Introduction     
%*****************************************************************************
\section{Introduction}

Dynamic wetting encompasses a class of interfacial phenomena relevant to technical applications in strategic areas ranging from materials science to energy research.
Critical applications such as coating, self-assembly, and microfluidic handling have motivated extensive efforts toward developing models for the dynamic wetting of solid surfaces.
Nonetheless, physical understanding and successful mathematical modeling of these phenomena still requires significant developments \cite{Lauga2006,Shikhmurzaev2011}. 
A comprehensive description of dynamic wetting must account for the coupling between molecular and hydrodynamic interactions taking place on length and time scales that extend over several orders of magnitude. 

Classical continuum-level descriptions rely on the Navier-Stokes equations for the  mathematical modeling of hydrodynamics, while microscopic interactions need to be coarse grained in order to render hydrodynamic boundary conditions at fluid-solid or fluid-fluid interfaces.
In order to model the physical coupling between microscopic interactions and hydrodynamics, boundary conditions for the wall velocity and the static/dynamic contact angles commonly take the form of algebraic equations \cite{Cox1998} or even partial differential equations \cite{Shikhmurzaev1994}. 
From a mathematical viewpoint, these boundary conditions are not just input parameter values, but rather involve additional model equations. 
Furthermore, the set of hydrodynamic equations must usually be supplemented with equations of state (EOS) and other constitutive relations that can make a continuum-level description cumbersome.

This type of classical continuum description, although widely employed, suffers from several limitations in dealing with nontrivial microscopic effects. 
In particular, the sharp-interface limit leads to well-known mathematical singularities and paradoxes; e.g. logarithmic stress singularities, the moving contact line paradox \cite{Dussan1979, deGennes1985}.
Such unphysical artifacts are readily removed  \cite{deGennes1985} by considering that fluid-fluid and fluid-solid interfaces are actually ``diffuse'' and thus have a finite thickness, determined by thermal diffusion and the range of action of molecular forces.
The interface thickness can be considered as a cutoff length \cite{Cox1998}, or characteristic microscopic scale of the physical system, below which the classical continuum model with a sharp interface is not valid.

A fundamental phenomenon observed at the microscopic scales is the presence of thin interfacial films (e.g. precursor films on hydrophilic surfaces \cite{deGennes1985, Book1}) within which dispersion forces (e.g. London--van der Waals forces) become significant. 
The mass and momentum fluxes through these interfacial films can produce nontrivial effects on the hydrodynamic behavior (e.g. effective slip velocity \cite{Shikhmurzaev1994, Koplik2001, Lauga2006}) and the equilibrium conditions prevailing in the fluid bulk \cite{Starov2009}.
It becomes therefore necessary to refine the level of modeling within the interfacial films in order to model dynamic wetting phenomena in many systems of practical interest (e.g. micro/nanoflows, colloids).

In this work, we propose a so-called ``mesoscopic'' model \cite{Shan1994,Shan2006,Benzi2009} embodying an augmented description of the solid-fluid interface.
Such a mesoscopic approach, based on the single-particle distribution, is convenient for modeling the net effects of microscopic interaction forces on the dynamics of macroscopic quantities (e.g. mass density, thermodynamic pressure, and fluid momenta). 
This strategy can be realized by using mean-field interaction potentials, or pseudo-potentials, that are scalar functions of macroscopic quantities (e.g. mass density).
Similarly to effective interaction potentials used in different forms of DLVO theory \cite{Book1, Starov2009}, pseudo-potentials are scaled by interaction parameters (e.g. attraction/repulsion Hamaker constants).
These interaction parameters determine an interfacial force per unit area, or disjoining pressure \cite{Deryagin1955,deGennes1985,Teletzke1988}, that can have both repulsive and attractive components varying as a function of the distance from the interface.
The disjoining pressure near the solid-fluid interface ultimately determines the contact angle at the apparent three-phase contact line \cite{Teletzke1988, Book1, Starov2009}.
Hence, the set of attraction/repulsion parameters employed determines implicitly the wetting properties of the solid.
Moreover, the attractive-repulsive character of the modeled interactions can lead to multiple local minima and maxima in the disjoining pressure.
These physical features give rise to nontrivial effects, such as the static hysteresis of the apparent contact angle on a (macroscopically) smooth surface \cite{Starov2009}.

This article is structured as follows. 
In Sec. II we formulate our mesoscopic model motivated by the physical insights presented above; the current implementation of this model is based on the numerical solution of a lattice Boltzmann (LB) equation.
A fundamental difference with previous LB models lies in the augmented treatment of the solid-fluid interface by means of a wall probability function and a pseudo-potential for solid-fluid interactions having two adjustable parameters that control the magnitude of attractive and repulsive interfacial forces.
The two-parameter model allows us to simulate surfaces that exhibit the same static properties, i.e. same static contact angles, yet different dynamic behavior, i.e. different static or dynamic contact angle hystereses.
In Sec. III we report numerical results for pressure-driven flows of a volatile fluid. 
We focus on the static contact angles of sessile droplets and advancing/receding contact angles of droplets under pressure-driven flow. 
The results reveal that the proposed model reproduces key features of real solid surfaces. 
In Sec. IV we conclude with a summary of the key results and outline potential directions for the presented approach.
%
%
%*****************************************************************************
% Section 2: LATTICE BOLTZMANN FORMULATION
%*****************************************************************************
%
\section{A mesoscopic model for hydrodynamics and interfacial phenomena}
The model we propose is based on the Boltzmann--BGK equation for a single-component fluid \cite{Shan1994,Benzi2009} 
\begin{equation}
\frac{\partial f}{\partial t}+ {\bf v} \cdot \nabla f
=-\frac{f-f^{eq}}{\tau}+ \frac{\delta f}{\delta t}.
\label{eq:BE--BGK}
\end{equation}
Here $f \equiv f({\bf x, v},t)$ is the single-particle probability distribution at position coordinate ${\bf x}$ and velocity coordinate ${\bf v}$ at a time instance $t$. 
The term ${\delta f}/{\delta t}$ on the right-hand side of Eq.~\ref{eq:BE--BGK} accounts for the action of both external and internal forces.
The single relaxation time $\tau$ is a model parameter that determines the kinematic viscosity, the mobility, and other transport coefficients.
The equilibrium distribution function is a Maxwell-Boltzmann distribution
\begin{equation}
 f^{eq}({\bf x},{\bf v},t)=\frac{\rho}{(2\pi\theta)^{\frac{D}{2}}}\exp\left[-\frac{({\bf v}-{\bf
 u})^{2}}{2\theta}\right],
\label{eq:feqc}
\end{equation}
where $D$ is the velocity space dimensionality, $\rho$ is the local mass density, and ${\bf u}$ is the fluid velocity. 
The distribution variance $\theta=k_B T/m$ is determined by the specific thermal energy ($T$: fluid temperature, $k_B$: Boltzmann constant, $m$: molecular mass) and defines the characteristic scale of diffusive processes.
For the studied single-component system we adopt a unit molecular mass, $m=1$, without loss of generality.
The LB model in this work is valid for isothermal flows ($T$=const.) where a classical hydrodynamic description involves the three leading moments of the distribution function:
\begin{equation}
{\bf M}^{(0)}({\bf x},t)=\int f({\bf x},{\bf v},t) \textrm{d}{\bf v}=\rho;
\label{eq:M0}
\end{equation}
\begin{equation}
{\bf M}^{(1)}({\bf x},t)=\int f({\bf x},{\bf v},t) {\bf v}\textrm{d}{\bf v}
=\rho{\bf u};
\label{eq:M1}
\end{equation}
\begin{equation}
{\bf M}^{(2)}({\bf x},t)=\int f({\bf x},{\bf v},t){\bf v}{\bf v}\textrm{d}{\bf v}=\rho{\bf u}{\bf u}+{\boldsymbol \sigma}.
\label{eq:M2}
\end{equation}
The second-order moment, ${\bf M}^{(2)}$, is the momentum flux tensor given by the sum of the macroscopic momentum flux, $\rho{\bf u}{\bf u}$, and the stress tensor, ${\boldsymbol \sigma}$;
 in a closed-form description,  ${\boldsymbol \sigma}$ is a functional of mass density and fluid velocity.
A comparable level of hydrodynamic description is attained with a second-order LB method \cite{Shan2006,colosqui2010} where the distribution function $f$ is a truncated Hermite expansion
\begin{eqnarray}
\label{eq:f2}
f({\bf x},{\bf v},t)&=& f^{M}({\bf v}) [{M}^{(0)}+{\textstyle \frac{1}{\theta}} {\bf M}^{(1)}:{\bf v}\\ \nonumber
                   &+&{\textstyle \frac{1}{2\theta^2}}({\bf M}^{(2)}-{M}^{(0)}\theta {\bf I}):({\bf v}{\bf v}-\theta {\bf I})].
\end{eqnarray}
Here $f^M=(2\pi\theta)^{-D/2} \exp(-{\bf v}^2/2\theta)$ is a Gaussian weight and ${\bf I}$ is the unit tensor.
High-order moments (${\bf M}^{(n)};~n>2$) are readily evaluated in terms of the three leading order moments (${\bf M}^{(n)};~n=0,2$) by using Eq.~\ref{eq:f2}.
A projection step \cite{Shan2006,colosqui2010} needs to be included in the LB algorithm in order to enforce the functional form in Eq.~\ref{eq:f2}.
\subsection{Lattice Implementation} 
\begin{figure}
\centerline{\includegraphics[angle=0,scale=0.28]{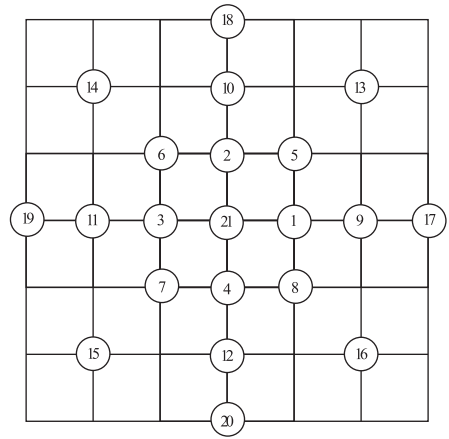}}
\caption{D2Q21 lattice cell. This lattice satisfies sixth-order moment isotropy.
Discrete gradient operators implemented on D2Q21 lattices retain isotropy up to the fifth-order spatial derivatives.}
\label{fig:D2Q21}
\end{figure}

The numerical algorithm evolves a finite set of distributions $f_i({\bf x},t) \equiv f({\bf x},{\bf v}_i,t)$ $(i=1,Q)$. 
The lattice velocities (${\bf v}_i;~i=1,Q$) are defined by a Gauss-Hermite quadrature rule \cite{Shan2006}.
The quadrature rule must fulfill a fourth-order algebraic degree of precision ($d\ge4$) to accurately approximate the isothermal flow solution in the continuum limit.
In the presence of large density gradients that can develop at an interface, a high-order rotational symmetry is highly desirable after discretization of velocity space. 
Such symmetry is critical to effectively retain the isotropy of high-order spatial derivatives required for the discrete approximation of interaction forces \cite{Shan2008}.
For the current implementation we adopt a D2Q21 lattice (dimensions $D=2$ and states $Q=21$)\cite{Shan2006}; abscissas and corresponding weights are reported in Tab.~\ref{tab:d2q21}.
%
%%%%%%%%%%%%%%%%%%%%%%%%%%%%%%%%%%%%%%%%%%%%%%%%%%%%%%%%%%%%%%%%%%%%%%%%%%%%%%%%%%%%%%%%%%%%%%%%%%%%%%%%%%%%%%%%%%%%%%%%%%%%%%%%
% Table D2Q21
%%%%%%%%%%%%%%%%%%%%%%%%%%%%%%%%%%%%%%%%%%%%%%%%%%%%%%%%%%%%%%%%%%%%%%%%%%%%%%%%%%%%%%%%%%%%%%%%%%%%%%%%%%%%%%%%%%%%%%%%%%%%%%%%
\begin{table}
\begin{center}
\begin{minipage}{7.5cm}
\begin{tabular}{@{}lcc@{}}
${\bf v}_i$ or ${\bf r}_i$ 
\footnote{$\theta=\sum_{i=1}^{21} w_i v_i^2 $ =  2/3 } & states & $w_i$ \\[0.5ex]
$(\pm 1,0)$,$(0,\pm1)$          & 1--4 &	$1/12$		\\[0.5ex]
$(\pm 1,\pm 1)$         	& 5--8 & 	$2/27$		\\[0.5ex]
$(\pm 2,0)$,$(\pm 2,0)$		& 9--12 &	$7/360$		\\[0.5ex]
$(\pm 2,\pm 2)$        	 	& 13--16 &	$1/432$		\\[0.5ex]
$(\pm 3,0)$,$(0,\pm 3)$        	& 17--20&	$1/1620$	\\[0.5ex]
$(0,0)$                 	& 21 &		$91/324$	\\[0.5ex]
\end{tabular}
\end{minipage}
\end{center}
\caption{Lattice D2Q21. Velocity abscissas (in lattice units) and Gauss-Hermite weights.}
\label{tab:d2q21}
\end{table}
%%%%%%%%%%%%%%%%%%%%%%%%%%%%%%%%%%%%%%%%%%%%%%%%%%%%%%%%%%%%%%%%%%%%%%%%%%%%%%%%%%%%%%%%%%%%%%%%%%%%%%%%%%%%%%%%%%%%%%%%%%%%%%%%
%
%
The D2Q21 lattice velocities are the integration points of a quadrature rule having seventh-order algebraic precision and satisfy moment isotropy up to the sixth-order \cite{Shan2006}.
This lattice retains isotropy of the fifth-order spatial derivatives in the discrete gradient operator (Eqs.~\ref{force_SC}--\ref{force_SC_FS})\cite{Shan2008}.
Thus, the D2Q21 lattice has comparatively better performance over low order lattices (e.g. D2Q9, D3Q17) in terms of reducing spurious currents and other numerical artifacts. 
The main disadvantage of using a high order lattice is that numerical simulation of even simple geometries can become computationally intensive.
A lattice cell (showed in Fig.~\ref{fig:D2Q21}) extends over six nodes; i.e. each node propagates information up to three lattice sites away in one time step.
%
%
%%%%%%%%%%%%%%%%%%%%%%%%%%%%%%%%%%%%%%%%%%%%%%%%%%%%%%%%%%%%%%%%%%%%%%%%%%%%%%%%%%%%%%%%%%%%%%%%%%%%%%%%%%%%%%%%%%%%%%%%%%%%%%%%
% Figure D2Q21
%%%%%%%%%%%%%%%%%%%%%%%%%%%%%%%%%%%%%%%%%%%%%%%%%%%%%%%%%%%%%%%%%%%%%%%%%%%%%%%%%%%%%%%%%%%%%%%%%%%%%%%%%%%%%%%%%%%%%%%%%%%%%%%%
%%%%%%%%%%%%%%%%%%%%%%%%%%%%%%%%%%%%%%%%%%%%%%%%%%%%%%%%%%%%%%%%%%%%%%%%%%%%%%%%%%%%%%%%%%%%%%%%%%%%%%%%%%%%%%%%%%%%%%%%%%%%%%%%
%
%
With the implementation in this work, properly resolving an interface or effectively rendering a boundary condition requires a minimum resolution of one lattice cell (i.e. six nodes).
For simulating nano- or microscale flows the computational cost of the present approach remains affordable by current computational resources \footnote{Simulations in this work required 1 to 4 hours of CPU time in a mid-size worksation (quad-core processor at 2.8Ghz and 8GB of RAM)} and relatively low when compared to molecular dynamics simulations.
Among other advantages of the proposed approach, the LB algorithm is particularly suitable for massive parallelization and/or acceleration using graphic processing units (GPUs) \cite{ZhaoGPU,*BernaschiGPU}.

\subsection{Non-ideal fluid behavior and interfacial phenomena} 
The mesoscopic description based on Eq.~\ref{eq:BE--BGK} employs the BGK ansatz to model the macroscopic effect of short-range fluid-fluid interactions.
All other interaction forces, responsible for non-ideal fluid behavior (e.g. non-ideal equation of state, phase separation) and interfacial phenomena (e.g. surface tension, disjoining pressure, partial wetting) are modeled by an approximation of the actual force term $\delta f/\delta t=- m^{-1} {\bf g} \cdot df/d{\bf v}$ in the kinetic transport equation; ${\bf g}$ is a body force that may also include external fields.
Among different possible approximations for  $\delta f/\delta t$ \cite{Shan2006,Benzi2009} we adopted the exact difference method introduced in \cite{Kupershtokh2009} because of its optimal numerical stability \cite{colosqui2012}.
Hence, the force term in Eq.~\ref{eq:BE--BGK} is $\delta f/\delta t=f^{eq}(\rho,{\bf u}^*)-f^{eq}(\rho,{\bf u})$ where the equilibrium distribution is computed using a ``shifted" velocity ${\bf u}^*={\bf u}+{\bf g} \Delta t$. 

For the class of LB models we employ \cite{Shan2006, Benzi2009}, the volumetric body force 
$\rho {\bf g}=\psi({\bf x},t) \nabla \int w(|{\bf r}|) \psi({\bf x}+{\bf r},t) \textrm{d}{\bf r}$ 
is determined by a spatial convolution of the pseudo-potentials 
$\psi$ with the Gaussian kernel $w(|{\bf r}|)=(2 \pi \kappa\theta)^{-2/D} exp(-|{\bf r}|^2/ 2\kappa\theta)$.
The characteristic length scale $\sqrt{\kappa \theta}$ of the interaction kernel determines the thickness of the resulting interfaces (in all our computations we use $\kappa=1$). 
The interaction force $\rho {\bf g}={\bf F}_{FF}+{\bf F}_{FS}$ contains a Fluid-Fluid component
\begin{equation}
{\bf F}_{FF}({\bf x},t) = \psi_{FF}({\bf x},t) \sum_{i=0}^{Q} w_i \psi_{FF}({\bf x}+{\bf r}_i,t) {\bf r}_i,
\label{force_SC}
\end{equation}
and a Fluid-Solid contribution
\begin{equation}
{\bf F}_{FS}({\bf x},t) = \rho({\bf x},t) \sum_{i=0}^{Q} w_i \psi_{FS}({\bf x}+{\bf r}_i,t) {\bf r}_i+{\bf \Delta F}_{S}({\bf x},t).
\label{force_SC_FS}
\end{equation}
The long-range fluid-solid interactions thus receive similar treatment to that  described in \cite{Benzi2009}, using different pseudo-potentials for cross interactions between species.
Hereafter, we refer to all those microscopic interactions that act beyond an atomic radius as {\it long-range}; in this work long-range forces are mainly associated to van der Waals interactions and ionic double layers. 

The last term, ${\bf \Delta F}_{S}$, introduces a momentum exchange between fluid and solid molecules attributed to short-range interactions (considered as probabilistic collision events) occurring in the region adjacent to the solid surface.
The discrete lattice directions, ${\bf r}_i$, and Gauss-Hermite weights, $w_i$, are the same as those employed for the D2Q21 lattice [see Tab.1].
All dimensional quantities in this work are reported in lattice units.

{\bf Fluid-Fluid interactions}. The Fluid-Fluid interactions are defined as:
\begin{equation}
\psi_{FF}({\bf x},t)= \sqrt{2 [\rho \theta-p_{EOS}(\rho,\theta)]},
\end{equation}
where the pressure $p_{EOS}$ is given by an equation of state (EOS).
In this work, we model a volatile fluid that can separate into a (stable) vapor and liquid phase at the studied temperature $T=m \theta/k_B$, while both phases exhibit ideal fluid behavior (i.e. $p \propto \rho$). 
The adopted EOS is given by a piecewise linear relation \cite{colosqui2012} 
\begin{equation}
p_{EOS}(\rho,\theta) = \left\{
\begin{array}{l l}
\rho   \theta_V  & \quad \mbox{$\rho \le \rho_1$}\\
p_1 + (\rho-\rho_1)   \theta_U & \quad \mbox{$\rho_1 < \rho \le \rho_2$}\\
p_2 + (\rho-\rho_2) \theta_L & \quad \mbox{$\rho > \rho_2$}\\
\end{array} \right.
\label{eq:EOS}
\end{equation}
where $\rho_1$ and $\rho_2$ are the endpoints of the unstable branch, $p_1=\rho_1 \theta_V$ and $p_2=\rho_1 \theta_V + (\rho_2-\rho_1) \theta_U$, 
and $\theta_V>0$, $\theta_U<0$, and $\theta_L>0$ are the slopes in the vapor, unstable, and liquid branches, respectively.
The parameters employed in the present work are: $\theta_V = 0.25 \theta$; $\theta_U = -0.25 \theta$;
$\theta_L =1.0 \theta$; $\rho_V=0.1$; $\rho_1=0.222$; $\rho_2=0.869$; and $\rho_L=1.0$.
Based on the continuum calculation of thermodynamic equilibrium, this parameter combination produces phase equilibrium at a density ratio $\rho_L/\rho_V=10$ and a compressibility ratio $\beta=\rho_L \theta_L/\rho_V \theta_V=1/40$.
For this system we report a surface tension $\gamma\simeq 0.09$ computed in simulations via force integration across a planar interface, as well as by measuring pressure differences across a circular interface [for a description of the procedure see \onlinecite{Kupershtokh2009, Sbragaglia2007}].

{\bf Fluid-Solid interactions}. The solid phase is treated on a similar footing with the liquid phase.
The proper choice of $\psi_{FS}$ can effectively model a disjoining pressure acting within a surface film of finite thickness.
The pseudo-potential employed for Fluid-Solid interactions has the general form
\begin{equation}
\psi_{FS}({\bf x},t)= G_R \bar{\psi}_R({\bf x}) + G_A \bar{\psi}_A({\bf x}).
\label{eq:pseudopotential}
\end{equation}
A procedure to determine the repulsive, $\bar{\psi}_R$, and attractive potentials, $\bar{\psi_A}$, for arbitrary surface geometries is described in the Appendix.
Once the pseudo-potentials are properly defined, the static contact angle $\theta_Y$ is modified by adjusting the attraction parameters $G_R$ and $G_A$.
In analogy with DLVO theory, the modeled Fluid-Solid interactions have a repulsive component $G_R \bar{\psi}_R$ (e.g. attributed to double-layer repulsion) and an attractive component $G_A \bar{\psi}_A$ (e.g. due to London--van der Waals forces).
The fundamental feature modeled in Eq.~\ref{eq:pseudopotential} is that the resulting surface forces reverse direction as the solid is approached.
The term in Eq.~\ref{force_SC_FS} that models short-range interactions (i.e. elastic collisions, Pauli repulsion), 
\begin{equation}
{\bf \Delta F_{S}}({\bf x},t)=\rho({\bf x},t) \phi_S^{\epsilon}({\bf x}) \Delta{\bf u}_{wall}({\bf x},t),
\label{eq:wall_shift}
\end{equation}
 produces a velocity shift $\Delta {\bf u}_{wall}= {\bf U}_{wall}-{\bf u}-{\bf F}_{FF} \Delta t$ in the fluid adjacent to the solid. 
The total momentum exchange between fluid and solid dynamically determines the hydrodynamic slip. 
In Eq.~\ref{eq:wall_shift} we consider that the solid surface can present a microscopic scale roughness that results in a diffuse fluid-solid interface for the present mesoscopic description.
Hence, the wall probability function $\phi_S({\bf x})$ takes finite values $0<\phi_S<1$ within the interfacial region where thin surface films develop. 
While $\phi_S=1$ inside the solid bulk, the collision probability vanishes ($\phi_S=0$) at a certain distance from the solid bulk.
In practice, the spatial variation of $\phi_S^{\epsilon}$ allows us to model the influence of microscale roughness on the effective slip \cite{Lauga2006}.
Once the wall function $\phi_S$ is defined (e.g. using the procedure in the Appendix), the exponent $\epsilon \ge 1$ in Eq.~\ref{eq:wall_shift} provides a means of incorporating different levels of microscale roughness via ``fine'' adjustments to the interface sharpness.
%
%
%*****************************************************************************
% Section 3: RESULTS
%*****************************************************************************
%
\section{Numerical Results}
\begin{figure}
\center
\includegraphics[angle=0,scale=0.4]{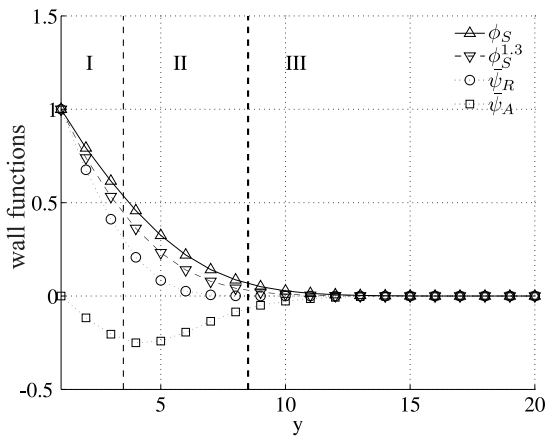}
\caption{
Wall functions determined before dynamic simulation: wall probability $\phi_S^\epsilon$ ($\epsilon=$ 1, 1.3); 
repulsive and attractive components of $\psi_{FS}=G_R \bar{\psi}_R+ G_A \bar{\psi}_A$; 
$y$ (in lattice units) is in the direction normal to a wall (bottom or top). 
The lines in the plots are interpolation curves fitting the computed node values  [see Appendix].
The value of the wall probability is used to determine three regions: 
(I) solid substrate ($\phi_S > 0.5$) where short-range (fluid-solid) interactions are dominant; 
(II) interfacial boundary layer ($0.5 \le \phi_S \le 0.01$) where surface forces are significant;
(III) hydrodynamic region or fluid bulk ($\phi_S < 0.01$) where surface forces become negligible.
}
\label{fig:solidwall}
\end{figure}

\begin{figure}
\includegraphics[angle=0,scale=0.4]{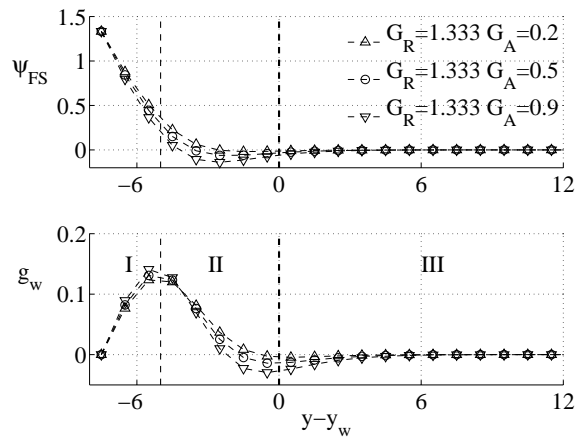}
\caption{
Fluid-solid potential (Top panel), $\psi_{FS}$, and volumetric force (Bottom panel), $g_w$, for interaction parameters within the range employed in simulations: $G_R=1.33$; $G_A=$ 0.2, 0.5, and 0.9.
The lines in the plots are interpolation curves fitting the computed node values  [see Appendix].
The $y$-axis is in the direction normal to a wall ($y$ coordinates are indicated in lattice units). 
The hydrodynamic region (I) begins at $y_w\simeq 8.5$ where $\phi_S(y_w)= 0.01$ (the opposite boundary is located at $y_w=L_y-7.5$). 
The normal component $g_w$ (given in lattice units) gives rise to a disjoining pressure $\Pi=-\rho g_w$ that dominates within the interfacial region (II).
}
\label{fig:wallforces}
\end{figure}

In this section, we present numerical results that demonstrate the capabilities of the method formulated in Sec. II and we discuss briefly the most relevant observations.
All simulations are performed employing small values of the relaxation time $\tau=$ 1.0--1.5; within this range the reported results present no significant dependence on $\tau$.
The only EOS employed and the corresponding equilibrium conditions are described in Sec. II.
The functions $\phi_S$, $\bar{\psi}_R$, and $\bar{\psi}_A$ that determine the solid location and all Fluid-Solid interactions are evaluated at initialization (i.e. before the actual dynamic simulation) using the numerical procedure described in the Appendix.
The wall functions employed for the present simulations are reported in Fig.~\ref{fig:solidwall}.
%
%%%%%%%%%%%%%%%%%%%%%%%%%%%%%%%%%%%%%%%%%%%%%%%%%%%%%%%%%%%%%%%%%%%%%%%%%%%%%%%%%%%%%%%%%%%%%%%%%%%%%%%%%%%%%%%%%%%%%%%%%%%%%%%%
% Figure Wall functions
%%%%%%%%%%%%%%%%%%%%%%%%%%%%%%%%%%%%%%%%%%%%%%%%%%%%%%%%%%%%%%%%%%%%%%%%%%%%%%%%%%%%%%%%%%%%%%%%%%%%%%%%%%%%%%%%%%%%%%%%%%%%%%%%
%%%%%%%%%%%%%%%%%%%%%%%%%%%%%%%%%%%%%%%%%%%%%%%%%%%%%%%%%%%%%%%%%%%%%%%%%%%%%%%%%%%%%%%%%%%%%%%%%%%%%%%%%%%%%%%%%%%%%%%%%%%%%%%%
%
We present in Fig.~\ref{fig:wallforces} the Fluid-Solid potential $\psi_{FS}$ and the volumetric force ${\bf g}_w$ for representative values of the interaction parameters.
For the simulated flat surfaces only the normal component of ${\bf g}_w$ is active, causing a disjoining pressure $\Pi=-\rho g_w$.
The curves in Fig.~\ref{fig:wallforces} illustrate the effects of scaling the attractive interaction tail of $\psi_{FS}$; this ultimately controls the surface wettability. 
%
%
%%%%%%%%%%%%%%%%%%%%%%%%%%%%%%%%%%%%%%%%%%%%%%%%%%%%%%%%%%%%%%%%%%%%%%%%%%%%%%%%%%%%%%%%%%%%%%%%%%%%%%%%%%%%%%%%%%%%%%%%%%%%%%%%
% Figure 3 Wall functions
%%%%%%%%%%%%%%%%%%%%%%%%%%%%%%%%%%%%%%%%%%%%%%%%%%%%%%%%%%%%%%%%%%%%%%%%%%%%%%%%%%%%%%%%%%%%%%%%%%%%%%%%%%%%%%%%%%%%%%%%%%%%%%%%
The modeled microscopic interactions give rise to three distinct spatial regions [see Figs.~\ref{fig:solidwall}--\ref{fig:wallforces}] where different effects dominate the dynamics.
The value of the wall probability ($0\le \phi_S \le 1$) determines the relative importance of fluid-solid interactions near the solid, and their effective absence within the fluid bulk.
We use characteristic values of the wall probability to define such regions:  
(I) solid substrate ($\phi_S>0.5$) where short-range interactions are dominant and a vapor fraction is adsorbed ($\rho/\rho_L \simeq 0.001$ for $G_R=3.0$, $\rho/\rho_L \simeq 0.03$ for $G_R=0.9$); 
(II) interfacial film ($ 0.5\le\phi_S\le 0.01$), or boundary layer,  where long-range interactions and the resulting disjoining pressure $\Pi=-\rho g_w$ dominate;
(III) fluid bulk ($\phi_S<0.01$), or hydrodynamic region, where surface forces are negligible and the Navier-Stokes equations are recovered.
In our simulations, the first node inside the hydrodynamic region ($\phi_S \le 0.01$) is located at 9 lattice units, or 1.5 lattice cells, from the outer boundary of the simulation domain.
The hydrodynamic velocity on the wall is observed at $y_w=8.5$, which corresponds to our limit ($\phi_S \simeq 0.01$) for the interfacial region ($y_w=L_y-7.5$ for the opposite wall with $L_y$ being the domain height).
%
%
%%%%%%%%%%%%%%%%%%%%%%%%%%%%%%%%%%%%%%%%%%%%%%%%%%%%%%%%%%%%%%%%%%%%%%%%%%%%
% Subsec:Pressure Driven Flow in capillaries
%%%%%%%%%%%%%%%%%%%%%%%%%%%%%%%%%%%%%%%%%%%%%%%%%%%%%%%%%%%%%%%%%%%%%%%%%%%%
%
\subsection{\label{sec:Poiseuille} Pressure-Driven Flow in capillaries}
Before studying problems that involve contact lines, we simulate pressure-driven flow of a volatile liquid ($\rho_L=1.0$) in a two-dimensional channel of dimensions $L_x\times L_y=10 \times 74$ with periodic boundary conditions in the $x-$direction.
A small pressure difference $\Delta_p=dp_x L_x$ is applied in the $x$-direction via a small body force $dp_x=4.0 \times 10^{-5}/\rho$, while we vary the interaction parameters that control surface wettability.
In order to model the effects of varying the microscopic-scale roughness we adjust the exponent of the wall probability function $\phi_S^\epsilon$; this exponent effectively adjusts the momentum exchange due to short-range repulsive interactions within the interfacial region (II) where long-range surface forces are active.

\begin{figure}
\center
\subfigure[~$G_R=1.33$, $G_A=0.2$, $\epsilon=1.0$]{\includegraphics[angle=0,scale=0.33]{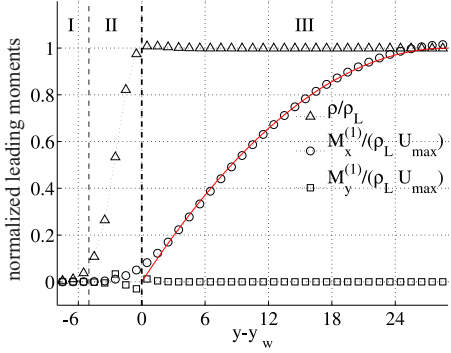}}\\
\subfigure[~$G_R=1.33$, $G_A=0.5$, $\epsilon=1.0$]{\includegraphics[angle=0,scale=0.33]{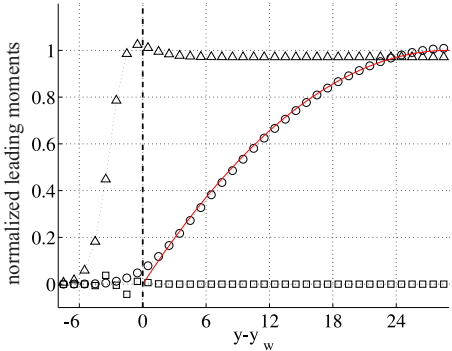}}\\
\subfigure[~$G_R=1.33$, $G_A=0.9$, $\epsilon=1.0$]{\includegraphics[angle=0,scale=0.33]{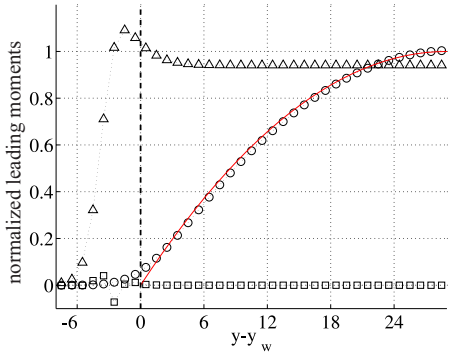}}
\caption{Pressure-driven flow on a microscopically ``rough'' surface ($\epsilon=1.0$) for different surface wettability conditions  ($G_R=1.33$, $G_A=$ 0.2, 0.5, and 0.9).
The (red) solid line indicates the analytical solution of incompressible Poiseuille flow, which is valid within the hydrodynamic region (III), for $\Delta p=4.0 \times 10^{-4}$ and using a no-slip boundary condition $u(y_w)=0$.
A surface film of varying density develops within the interfacial region (II).}
\label{fig:GR133}
\end{figure}

\begin{figure}
\center
\subfigure[~$G_R=1.33$, $G_A=0.2$, $\epsilon=1.3$]{\includegraphics[angle=0,scale=0.33]{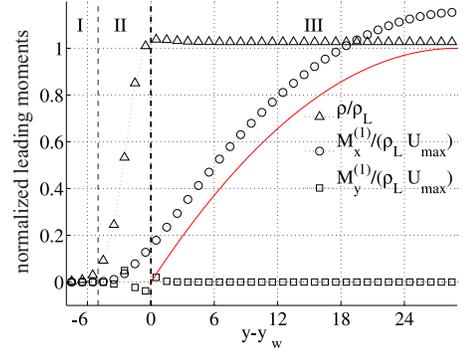}}\\
\subfigure[~$G_R=1.33$, $G_A=0.5$, $\epsilon=1.3$]{\includegraphics[angle=0,scale=0.33]{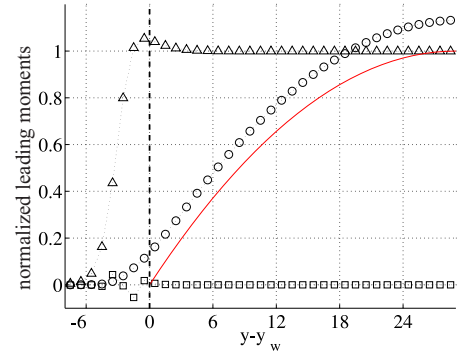}}\\
\subfigure[~$G_R=1.33$, $G_A=0.9$, $\epsilon=1.3$]{\includegraphics[angle=0,scale=0.33]{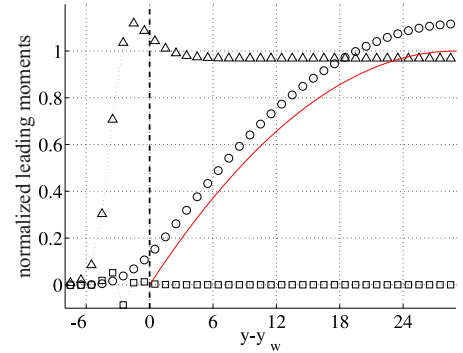}}
\caption{Pressure-driven flow on a microscopically ``smooth'' surface ($\epsilon=1.3$) for different surface wettability conditions ($G_R=1.33$, $G_A=$ 0.2, 0.5, and 0.9).
The (red) solid line indicates the analytical solution of incompressible Poiseuille flow, which is valid within the hydrodynamic region (III), for $\Delta p=4.0 \times 10^{-4}$ and using a no-slip boundary condition $u(y_w)=0$.}
\label{fig:GR133_wfe130}
\end{figure}

The results in Figs.~\ref{fig:GR133}--\ref{fig:GR133_wfe130} report mass density and fluid momentum profiles for three sets of interaction parameters ($G_R=1.33$, $G_A=$ 0.2, 0.5, 0.9).
We compare two values of the wall function exponent: 
$\epsilon=$ 1.0 to model a ``rough'' surface where microscopic roughness and long-range interactions have the same characteristic length; and
$\epsilon=$ 1.3 to model a ``smooth'' surface where the microscopic roughness is smaller than the range of action of long-range surface forces.
The density profile in Figs.~\ref{fig:GR133}--\ref{fig:GR133_wfe130} shows an interfacial film that develops between the hydrodynamic region and the solid substrate.
We observe that the thickness and density of the surface films are adjusted by varying the interaction parameters, $G_R$ and $G_A$, that control the Fluid-Solid, $\psi_{FS}$, and Fluid-Fluid, $\psi_{FF}$, pseudo-potentials.
It follows that the modeled physico-chemical properties of the fluid and solid phases determine the hydrodynamic slip.
The amount of hydrodynamic slip is determined by comparing against the analytical solution of incompressible Poiseuille flow with no-slip velocity applied at $y_w$.
%
%
%%%%%%%%%%%%%%%%%%%%%%%%%%%%%%%%%%%%%%%%%%%%%%%%
% Figure 4 Poiseuille flow
%%%%%%%%%%%%%%%%%%%%%%%%%%%%%%%%%%%%%%%%%%%%%%%%
%%%%%%%%%%%%%%%%%%%%%%%%%%%%%%%%%%%%%%%%%%%%%%%%%%%%%%%%%%%%%%%%%%%%%%%%%%%%%%%%%%%%
%
%%%%%%%%%%%%%%%%%%%%%%%%%%%%%%%%%%%%%%%%%%%%%%
% Figure Poiseuille flow
%%%%%%%%%%%%%%%%%%%%%%%%%%%%%%%%%%%%%%%%%%%%%%
%%%%%%%%%%%%%%%%%%%%%%%%%%%%%%%%%%%%%%%%%%%%%%%%%%%%%%%%%%%%%%%%%%%%%%%%%%%%%%%%%%%%
In agreement with molecular dynamic simulations of similar pressure-driven flows \cite{Koplik2001}, the amount of effective slip is rather independent of the studied flow conditions.
As seen in Figs.~\ref{fig:GR133}--\ref{fig:GR133_wfe130}, there is no clear correlation between the hydrodynamic slip and the surface wettability, determined by the attraction parameter ($G_A=$ 0.2--0.9) when the repulsion  parameter is fixed ($G_R=1.33$).
The scale of the (modeled) micro-roughness, controlled by $\epsilon$, has significant effects on the hydrodynamic slip.
A microscopically ``rough'' surface ($\epsilon=1.0$) presents no effective slip [Fig.~\ref{fig:GR133}], regardless of its wettability, while a ``smooth'' surface exhibits significant slip [Fig.~\ref{fig:GR133_wfe130}] even for hydrophilic conditions produced by a high attraction parameter ($G_A=0.9$).
We remark that these qualitative physical features modeled with the present approach are reported in a large number of experimental studies \cite{Lauga2006}.
%
%
%%%%%%%%%%%%%%%%%%%%%%%%%%%%%%%%%%%%%%%%%%%%%%%%%%%%%%%%%%%%%%%%%%%%%%%%%%%%
% Subsec:Equilibrium contact angle
%%%%%%%%%%%%%%%%%%%%%%%%%%%%%%%%%%%%%%%%%%%%%%%%%%%%%%%%%%%%%%%%%%%%%%%%%%%%
%
\subsection{\label{sec:static} Static wetting and equilibrium contact angle}
\begin{figure}
\center
\subfigure[]{\includegraphics[angle=0,scale=0.4]{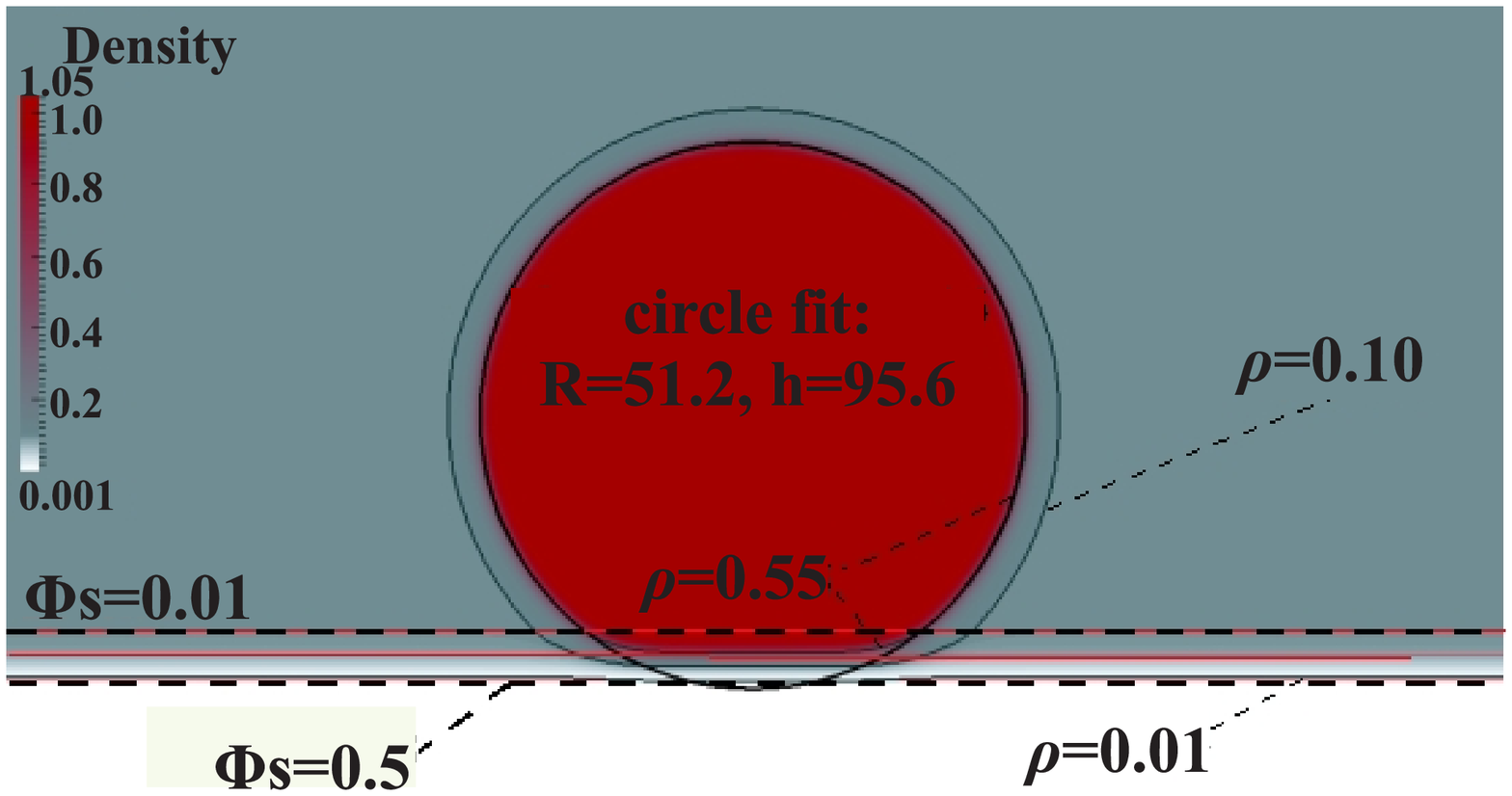}}\\
\subfigure[]{\includegraphics[angle=0,scale=0.4]{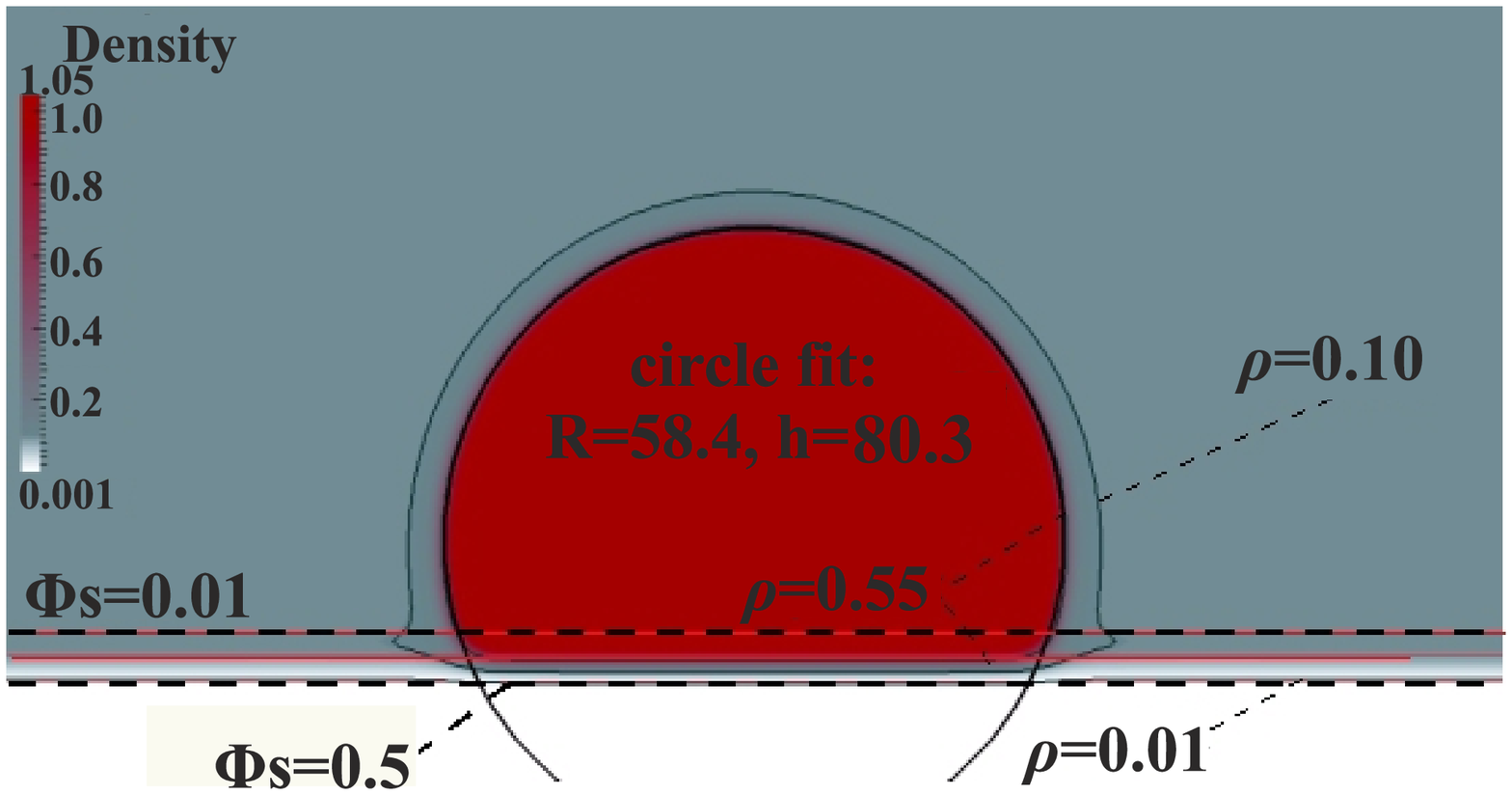}}\\
\subfigure[]{\includegraphics[angle=0,scale=0.4]{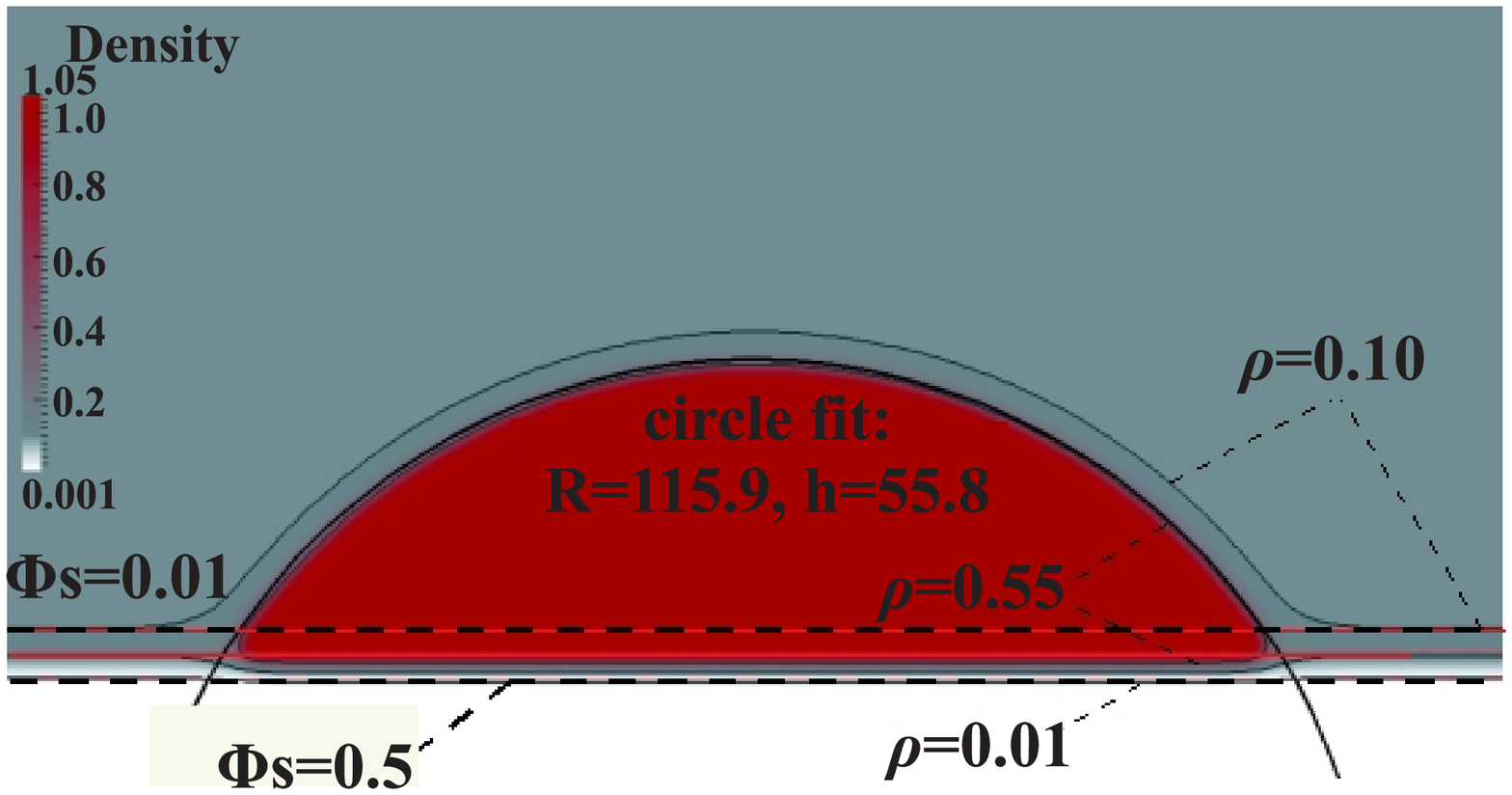}}\\
\subfigure[]{\includegraphics[angle=0,scale=0.4]{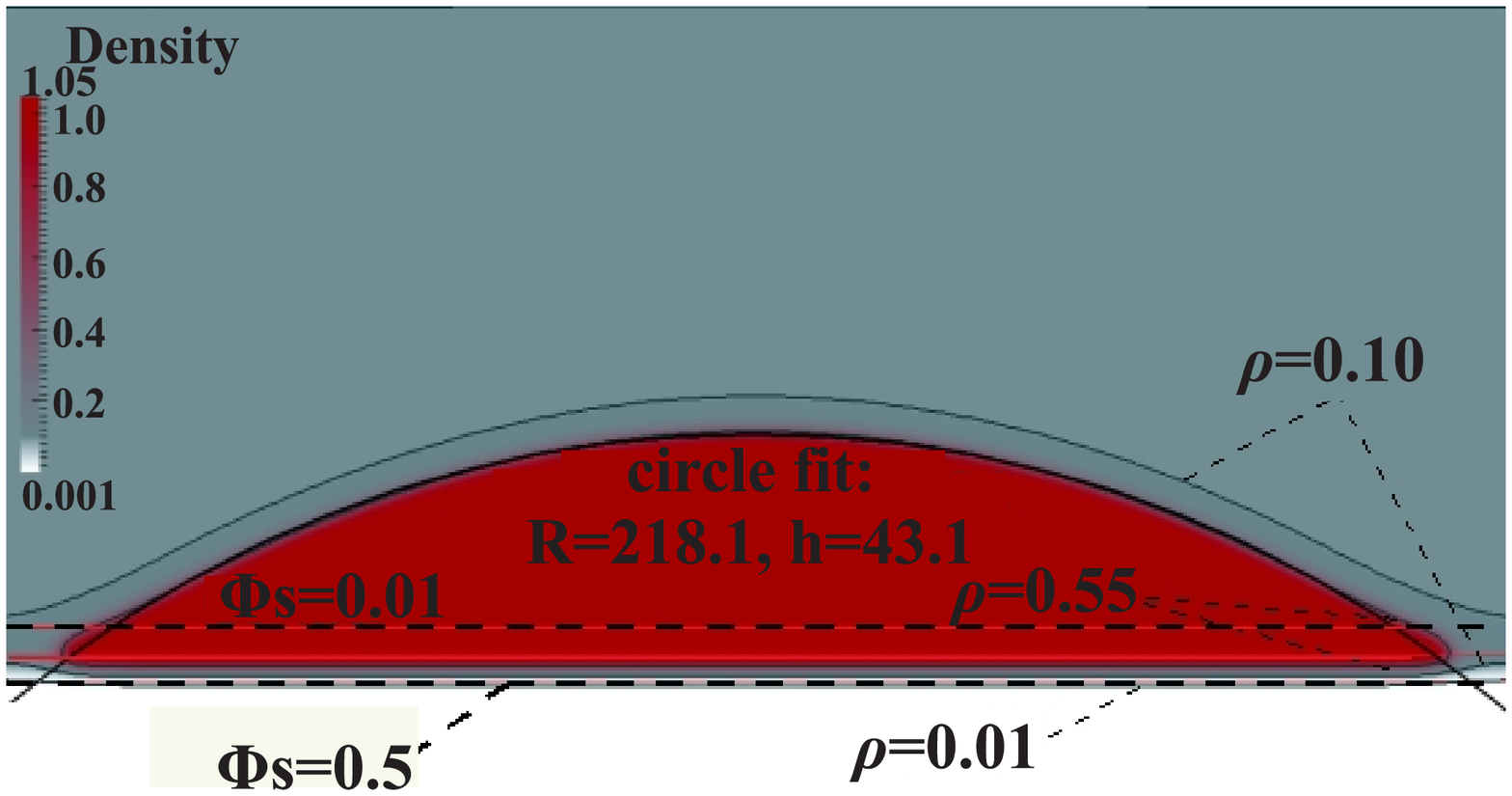}}
\caption{Sessile circular droplet at equilibrium.  
The equilibrium contact angle $\theta_Y=\mathrm{acos}(1-h/R)$ is computed via circle fitting (solid black line); $R$ is the circle radius, and $h$ indicates the droplet height (measured from the outer boundary of the simulation domain).
Contour lines (solid) for the fluid density, $\rho=0.01, 0.1, 0.55$, indicate a gradual growth of the interfacial film thickness as the attraction parameter $G_A$ increases.
Contour lines (dashed) for the solid probability function, $\rho_w=0.1, 0.5$, contain the interfacial region where a disjoining pressure is active.
All cases correspond to $G_R=1.33$ and $R_{0}=50$.
(a) $G_A=0.0$, $\theta_Y=150^\circ$. 
(b) $G_A=0.2$, $\theta_Y=112^\circ$.
(c) $G_A=0.5$, $\theta_Y=58.7^\circ$. 
(d) $G_A=0.7$, $\theta_Y=36.7^\circ$.}
\label{fig:sessile_drop}
\end{figure}

\begin{figure}
\center
\subfigure[]{\includegraphics[angle=0,scale=0.4]{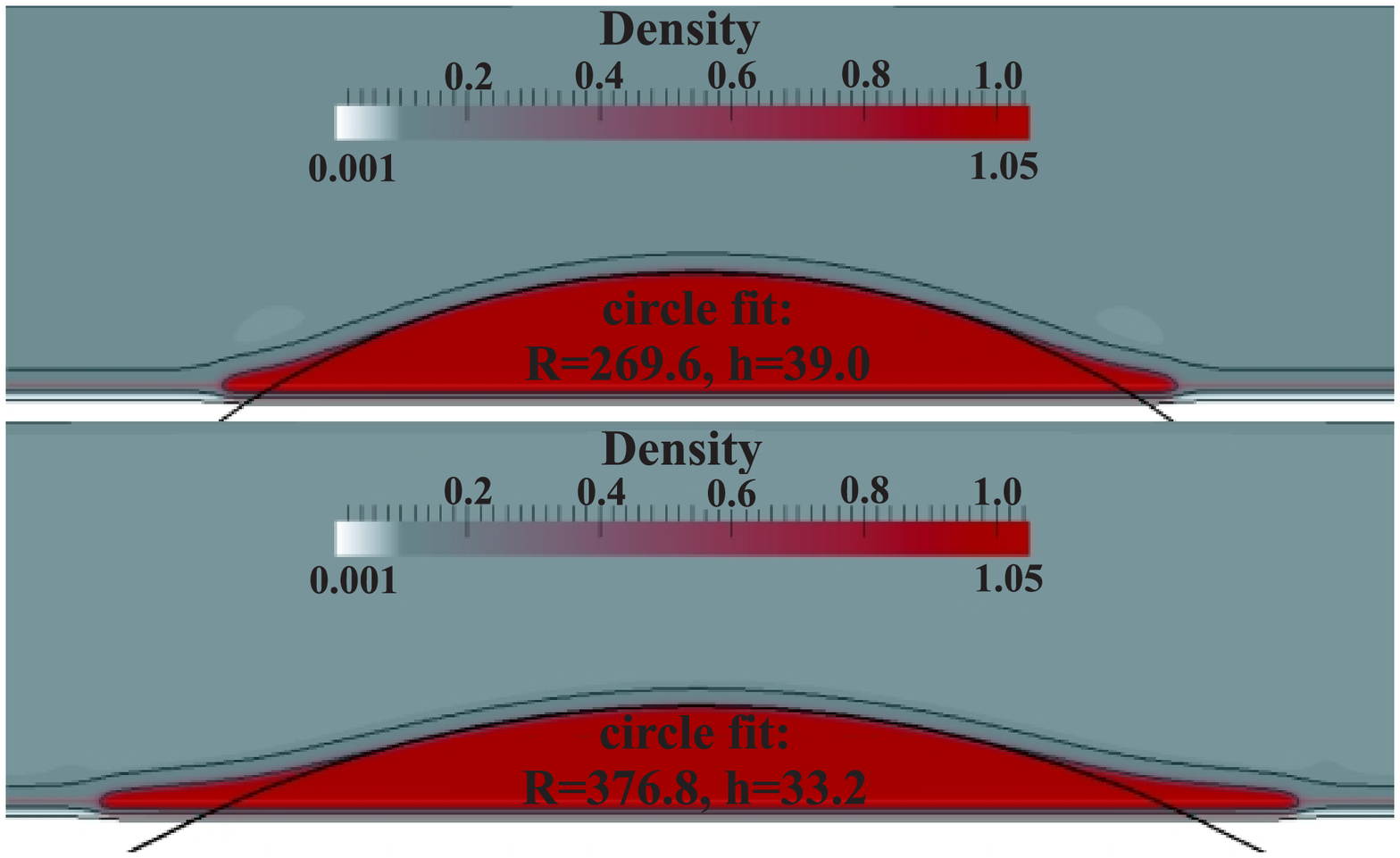}}\\
\subfigure[]{\includegraphics[angle=0,scale=0.4]{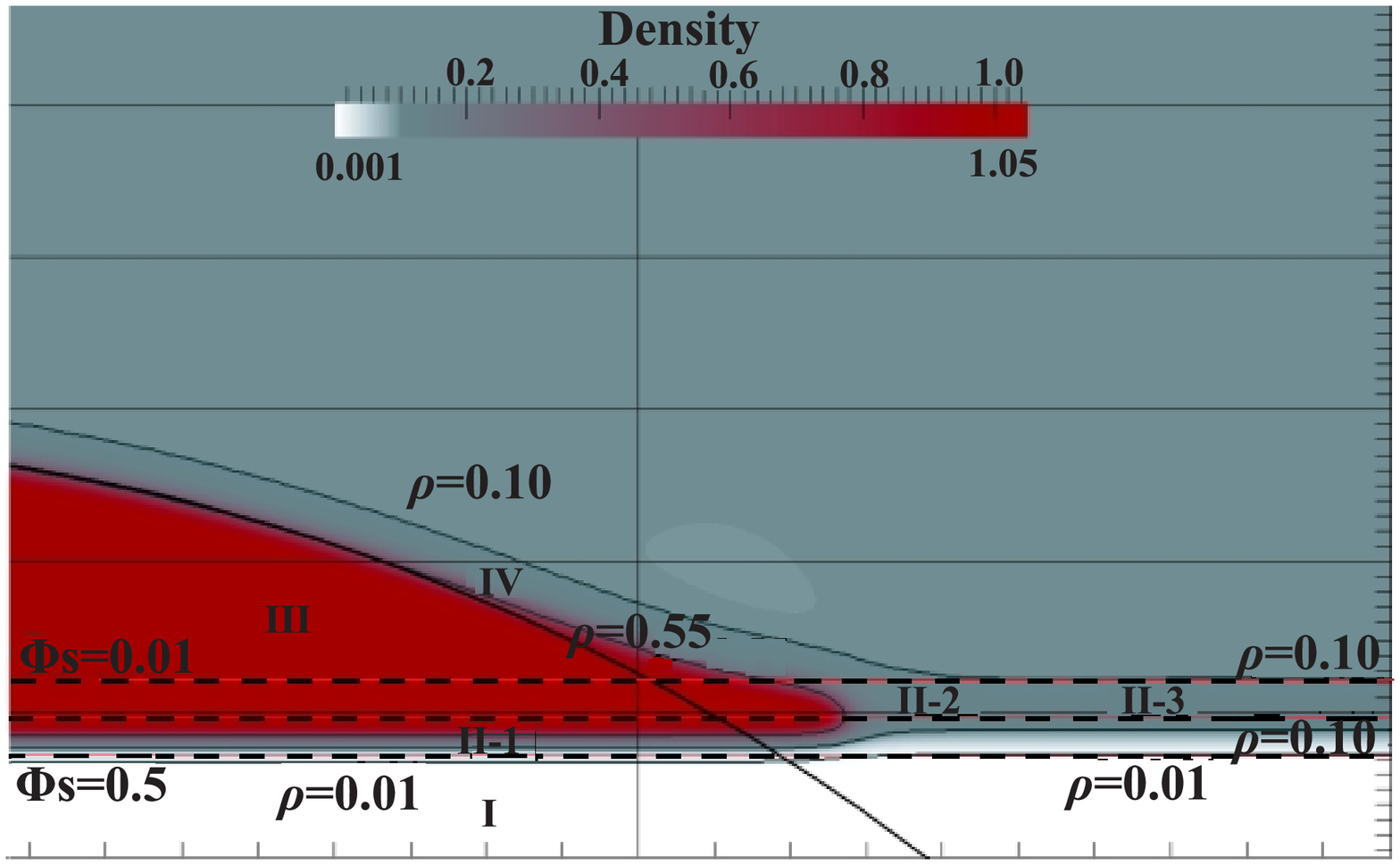}}
\caption{Sessile circular droplet on a hydrophilic surface.  
Circle fitting and contour line description is as in Fig.~\ref{fig:sessile_drop} for $G_R=1.33$ and $R_{0}=50$.
(a) Top panel: $G_A=0.8$, $\theta_Y=30.9^\circ$; bottom panel: $G_A=0.9$, $\theta_Y=24.0^\circ$.
(b) Density field in the vicinity of the (apparent) three-phase contact line [see \cite{Starov2009}]. 
I: Solid bulk, where a small vapor fraction ($\rho<0.001$) is adsorbed;
II-1: interfacial film at the liquid–solid interface;
II-2: overlap region between liquid-vapor and liquid-solid interfaces;
II-3: thin interfacial film at the vapor-solid interface.
III: liquid bulk.
IV: liquid-vapor interfacial region.
Interfacial forces (e.g. disjoining pressure) in regions II-(1--3) are determined by both interaction pseudo-potentials $\psi_{FS}$ and $\psi_{FF}$.}
\label{fig:sessile_drop2}
\end{figure}

We simulate a two-dimensional drop on a flat surface in order to quantify the surface wettability when varying the surface interaction potentials through the parameters $G_R$ and $G_A$. 
The simulation domain size is $L_x \times L_y=650 \times 150$ lattice units and the drop volume per unit width is $V_d=\pi R_0^2$; the results present no significant dependence on the droplet volumes employed ($R_0=$ 30--80) within the studied range of contact angles $20^\circ \le \theta_Y \le 160^\circ$.
The equilibrium contact angle is determined by numerically fitting a circle of radius $R$ [see Figs.~\ref{fig:sessile_drop}--\ref{fig:sessile_drop2}] to the vapor-liquid interface which is defined by the contour line for $\rho=(\rho_L+\rho_V)/2$.
The circle fitting by least squares is confined to the hydrodynamic region ($\phi_S>0.01$), above the interfacial film, where surface forces are negligible and constant curvature of the droplet is to be expected \cite{Starov2009}.
The reported equilibrium contact angle $\theta_Y=\mathrm{acos}(1-h/R)$ is evaluated using the circle radius $R$ and droplet height $h$ (i.e. distance between the apex and the bottom liquid-vapor interface) [see Figs.~\ref{fig:sessile_drop}--\ref{fig:sessile_drop2}].
The {\em apparent} contact angle, which we do not report in this work, should be measured above the interfacial surface film where $y=y_w$.
%
%%%%%%%%%%%%%%%%%%%%%%%%%%%%%%%%%%%%%%%%%%%%%%%%%%%%%%%%%%%%%%%%%%%%%%%%%%%%%%%%%%%%%%%%%%%%%%%%%%%%%%%%%%%%%%%%%%%%%%%%%%%%%%%%
% Figure 6: Sessile droplet low-to-moderate wettability
%%%%%%%%%%%%%%%%%%%%%%%%%%%%%%%%%%%%%%%%%%%%%%%%%%%%%%%%%%%%%%%%%%%%%%%%%%%%%%%%%%%%%%%%%%%%%%%%%%%%%%%%%%%%%%%%%%%%%%%%%%%%%%%%
%%%%%%%%%%%%%%%%%%%%%%%%%%%%%%%%%%%%%%%%%%%%%%%%%%%%%%%%%%%%%%%%%%%%%%%%%%%%%%%%%%%%%%%%%%%%%%%%%%%%%%%%%%%%%%%%%%%%%%%%%%%%%%%%
%

\begin{figure}
\center
\subfigure[]{\includegraphics[angle=0,scale=0.4]{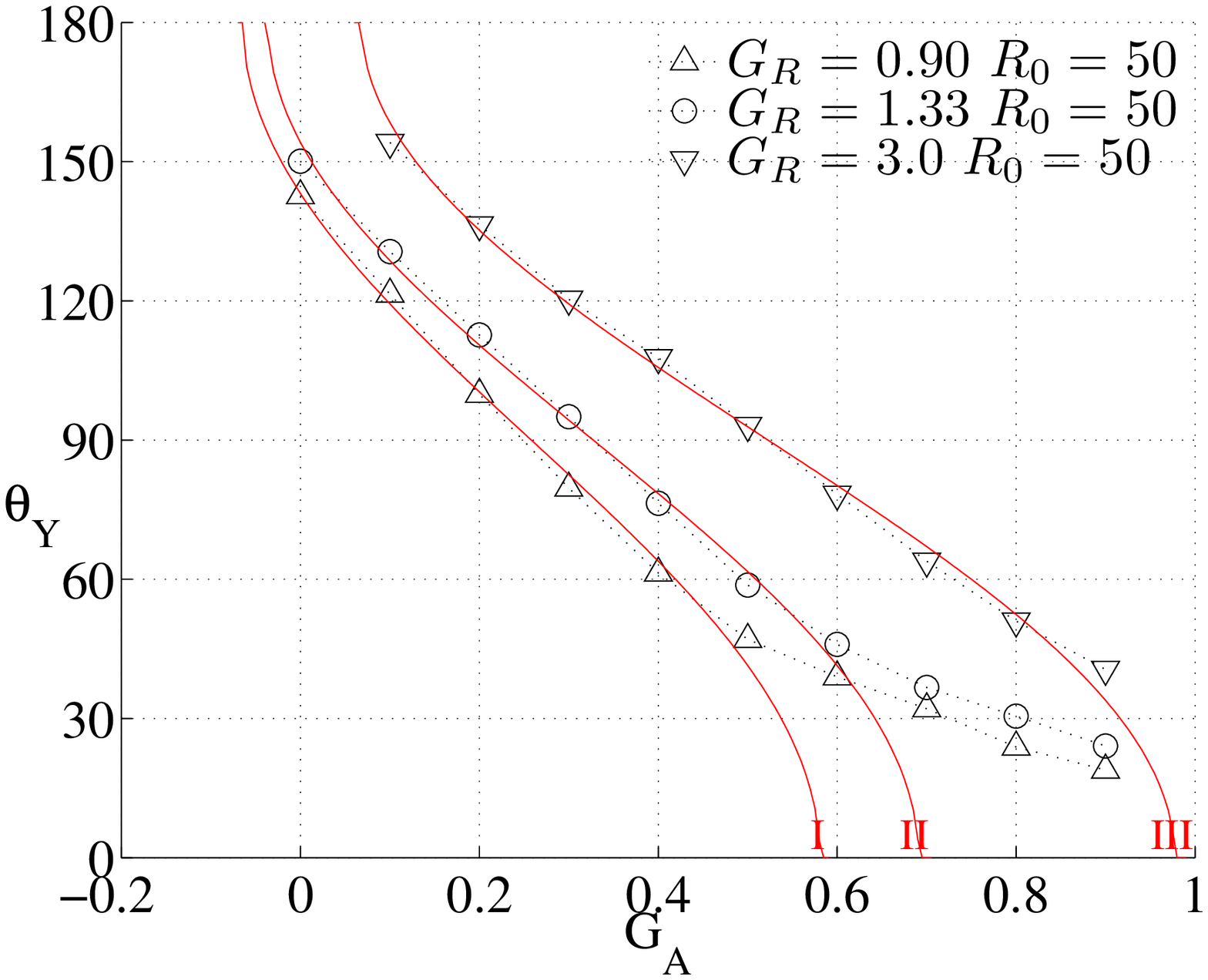}}
\subfigure[]{\includegraphics[angle=0,scale=0.4]{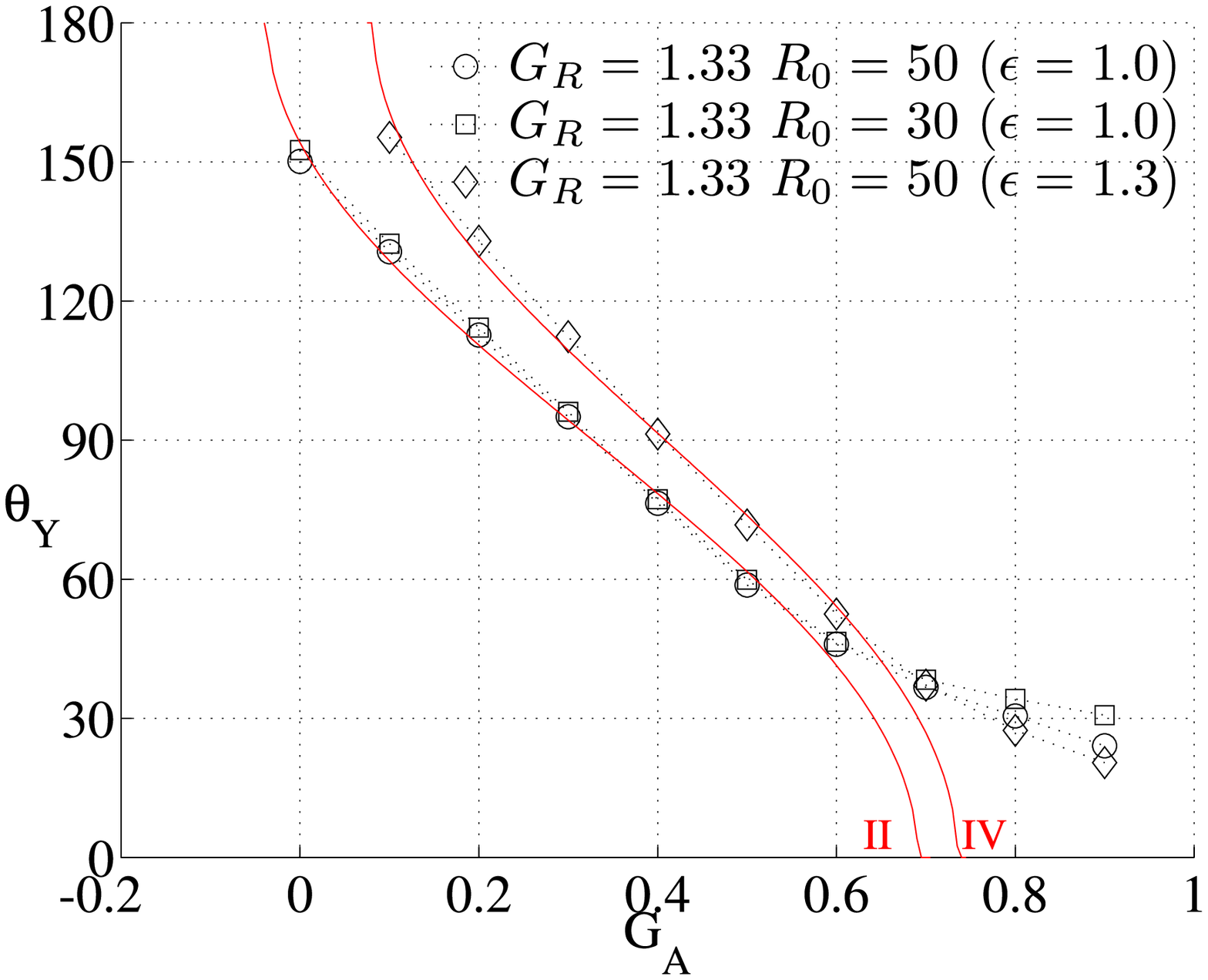}}
\caption{Equilibrium contact angle $\theta_Y$ for a two-dimensional droplet on a flat surface versus attraction parameter $G_A$.
The reference radius $R_0$ determines the droplet volume (per unit width) $V_d=\pi R_0^2$.
The same equilibrium contact angle can be obtained using infinite combinations of wall interaction parameters $G_R$ and $G_A$.
Solid lines are estimations by an augmented Young-Laplace equation $ \cos \theta_Y \approx (E_{0}+k_A G_A)/\gamma$:
(I) $E_0=-0.64$, $k_A=2.5$; (II) $E_0=-0.87$, $k_A=2.2$; (III) $E_0=-1.92$, $k_A=1.76$; (IV) $E_0=-1.0$, $k_A=3.06$; in all cases $\gamma=0.09$.
Deviations from a linear approximation for the surface energy excess $\Delta E= k_A G_A$ are observed for $\theta_Y<45^\circ$.
}
\label{fig:theta_GA}
\end{figure}

The density field reported in Figs.~\ref{fig:sessile_drop}--\ref{fig:sessile_drop2} exhibits fundamental physical features observed in previous work accounting for the presence of van der Waals forces [see \cite{Deryagin1955, deGennes1985, Teletzke1988, Starov2009}].
The most relevant feature observed in our simulations is the development of interfacial films between the sessile droplet and the solid substrate; in Fig.~\ref{fig:sessile_drop2} we identify the overlapping boundary regions analytically studied in \cite{Starov2009}.
As expected, interfacial forces adjusted via $G_A$ and $G_R$ determine the position of the vapor-liquid interface ($\rho=0.55$), at which mechanical equilibrium is attained, as well as the equilibrium contact angle.
%
%%%%%%%%%%%%%%%%%%%%%%%%%%%%%%%%%%%%%%%%%%%%%%%%%%%%%%%%%%%%%%%%%%%%%%%%%%%%%%%%%%%%%%%%%%%%%%%%%%%%%%%%%%%%%%%%%%%%%%%%%%%%%%%%
% Figure 7: Sessile droplet high surface wettability
%%%%%%%%%%%%%%%%%%%%%%%%%%%%%%%%%%%%%%%%%%%%%%%%%%%%%%%%%%%%%%%%%%%%%%%%%%%%%%%%%%%%%%%%%%%%%%%%%%%%%%%%%%%%%%%%%%%%%%%%%%%%%%%%
%%%%%%%%%%%%%%%%%%%%%%%%%%%%%%%%%%%%%%%%%%%%%%%%%%%%%%%%%%%%%%%%%%%%%%%%%%%%%%%%%%%%%%%%%%%%%%%%%%%%%%%%%%%%%%%%%%%%%%%%%%%%%%%%
%
The equilibrium contact angle is reported in Fig.~\ref{fig:theta_GA} as a function of the attraction parameter $G_A$ for three different values of the repulsive parameter $G_R$ and the modeled micro-roughness effects studied in Sec.\ref{sec:Poiseuille}.
Employing the augmented Young-Laplace equation \cite{Book1, Teletzke1988, Starov2009} the equilibrium contact angle is determined by
\begin{eqnarray}
\label{eq:augmentedYL}
 \cos \theta_Y &=& 1 + \frac{1}{\gamma} \int_{0}^{\infty} \Pi(y) dy\\ \nonumber 
                     &=& \frac{1}{\gamma} \left[E_{0}(G_R) + k_A G_A \right]+{\cal O}(G_A^2)
\end{eqnarray}
after adopting a linear approximation for the surface energy difference ``excess'' $\Delta E = E_{0}(G_R)+k_A G_A$.
The linear approximation in Eq.~\ref{eq:augmentedYL} fits the numerical data in Fig.~\ref{fig:theta_GA}(a--b) for moderate-to-large values of the contact angle.

Similar behavior of the contact angle variation with the surface energy excess, and an apparent saturation of the minimum contact angle attained, has been reported by an alternative approach based on Cahn-Hilliard models for the surface free-energy \cite{Pooley2009,Mognetti2010}.
The speculated reason for the apparent saturation of the minimum contact angle  is geometric effects \cite{Mognetti2010};
because the size of the droplet is comparatively small with respect to the surface film, at low but finite contact angles the droplet transitions into a thick film and the circle fitting procedure becomes inaccurate.
Nevertheless, the reported behavior at low contact angles presented no significant dependence on the droplet sizes employed ($R_0=$ 30--80) in our simulations.
Compressibility effects can also contribute to the observed deviations from a linear increase in the surface energy $\Delta E$ with respect to $G_A$ and the apparent saturation to a minimum contact angle attained.
%
%
%%%%%%%%%%%%%%%%%%%%%%%%%%%%%%%%%%%%%%%%%%%%%%%%%%%%%%%%%%%%%%%%%%%%%%%%%%%%%%%%%%%%%%%%%%%%%%%%%%%%%%%%%%%%%%%%%%%%%%%%%%%%%%%%
% Figure 8: theta vs GA
%%%%%%%%%%%%%%%%%%%%%%%%%%%%%%%%%%%%%%%%%%%%%%%%%%%%%%%%%%%%%%%%%%%%%%%%%%%%%%%%%%%%%%%%%%%%%%%%%%%%%%%%%%%%%%%%%%%%%%%%%%%%%%%%
%%%%%%%%%%%%%%%%%%%%%%%%%%%%%%%%%%%%%%%%%%%%%%%%%%%%%%%%%%%%%%%%%%%%%%%%%%%%%%%%%%%%%%%%%%%%%%%%%%%%%%%%%%%%%%%%%%%%%%%%%%%%%%%%
%
A key observation in this section is that infinite combinations of interaction parameters ($G_R$,$G_A$) can produce the same equilibrium contact angle. 
In the following subsection we study dynamic wetting properties of surfaces that exhibit the same static contact angle, attained by different combinations of interaction parameters.
%
%
%%%%%%%%%%%%%%%%%%%%%%%%%%%%%%%%%%%%%%%%%%%%%%%%%%%%%%%%%%%%%%%%%%%%%%
% Subsection: Contact angle dynamics
%%%%%%%%%%%%%%%%%%%%%%%%%%%%%%%%%%%%%%%%%%%%%%%%%%%%%%%%%%%%%%%%%%%%%%
%
\subsection{Contact Angle Dynamics}
\begin{figure}
\center
\includegraphics[angle=0,scale=0.4]{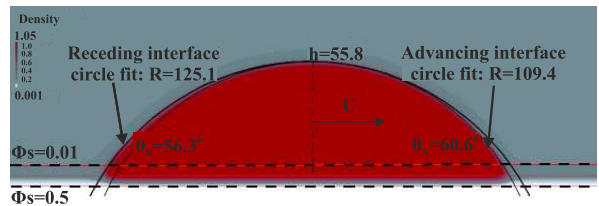}
\caption{Dynamic contact angles for a moving droplet.
Advancing/receding contact angles, $\theta_A$ and $\theta_R$, are determined by fitting two circles, of  radii $R$, to the vapor-liquid interface from the apex, at height $h$, to the corresponding advancing/receding contact lines.
The illustrated case corresponds to $Ca=0.006$ using interaction parameters $G_R=1.33$ and $G_A=0.5$, which produce an equilibrium contact angle $\theta_Y \simeq 58^\circ$.}
\label{fig:hysteresis}
\end{figure}

After characterizing static wetting properties of the modeled surfaces we investigate dynamic wetting conditions by applying a pressure difference that causes a flow in the $x$-direction and motion of the droplet; these simulations are performed for the same domain as in Sec.~\ref{sec:static}.
The same static contact angle $\theta_Y \simeq 58^\circ$ is obtained with three sets of interaction parameters: ($G_R=1.33$, $G_A=0.5$) and ($G_R=0.9$, $G_A=0.42$) with $\epsilon=1.0$ for the modeled micro-roughness that exhibited no hydrodynamic slip in Sec.~\ref{sec:Poiseuille}; and ($G_R=1.33$, $G_A=0.57$) with $\epsilon=1.3$ for the surface that exhibited hydrodynamic slip in Sec.~\ref{sec:Poiseuille}.
Contact angle values are numerically measured via circle fitting within the hydrodynamic region ($\phi_S < 0.01$) with a procedure similar to that employed for the static contact angle.
As showed in Fig.~\ref{fig:hysteresis}, advancing and receding angles are evaluated with the radii of the two circles that fit the rear and front interfaces connecting at height $h$.
%
%
%%%%%%%%%%%%%%%%%%%%%%%%%%%%%%%%%%%%%%%%%%%%%%%%%%%%%%%%%%%%%%%%%%%%%%%%%%%%%%%%%%%%%%%%%%%%%%%%%%%%%%%%%%%%%%%%%%%%%%%%%%%%%%%%
% Figure 9: Circle fitting dynamic 
%%%%%%%%%%%%%%%%%%%%%%%%%%%%%%%%%%%%%%%%%%%%%%%%%%%%%%%%%%%%%%%%%%%%%%%%%%%%%%%%%%%%%%%%%%%%%%%%%%%%%%%%%%%%%%%%%%%%%%%%%%%%%%%%
%%%%%%%%%%%%%%%%%%%%%%%%%%%%%%%%%%%%%%%%%%%%%%%%%%%%%%%%%%%%%%%%%%%%%%%%%%%%%%%%%%%%%%%%%%%%%%%%%%%%%%%%%%%%%%%%%%%%%%%%%%%%%%%%
%

\begin{figure}
\center
\includegraphics[angle=0,scale=0.4]{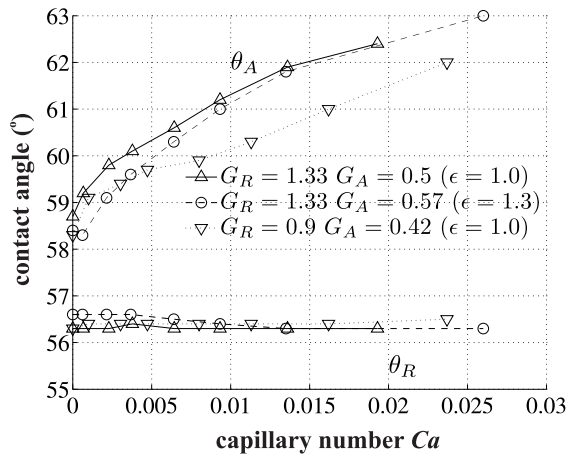}
\caption{Advancing and receding contact angles versus capillary number $Ca=U \mu_L/ \gamma$ (for the definition of the appropriate $U$ see text).
All sets of interaction parameters result in the same value for the equilibrium angle $\theta_Y\simeq 58^\circ$
but different contact angle dynamics.
The functional form of the disjoining pressure produces a finite static hysteresis of the contact angles.
}
\label{fig:Ca_qAqR}
\end{figure}

The applied pressure difference ($\Delta p=$ 0.175--5.25 $\times 10^{-4}$) causes a linear increase in the mean droplet velocity $U=$ 0--0.01;  the advancing and receding contact lines move with a velocity approximately equal to the computed mean speed $U$.
Thus we determine a capillary number $Ca=U \mu_L/ \gamma$ which we employ in Fig.~\ref{fig:Ca_qAqR} to report the measured dynamic contact angles.
The key observations from Fig.~\ref{fig:Ca_qAqR} are the following.
(1) The model produces a small but finite contact angle hysteresis in static conditions ($Ca\to 0$) for a macroscopically smooth surface.
We remark that we only have considered roughness at a microscale comparable the range of action of surface forces, which we modeled with a wall probability $\phi_S^\epsilon$.
(2) The static hysteresis of the contact angles is produced by the functional shape of the disjoining pressure, as predicted in \cite{Starov2009}, and shows no significant dependence on the modeled microscale roughness.
(3) Advancing and receding contact angles vary at different rates; while the advancing contact angle $\theta_A$ increases with $Ca$, the receding contact angle $\theta_R$ is weakly affected.
This last observation indicates the influence of the flow structure in the vicinity of a moving contact line and is further discussed in the next Section.
%
%
%%%%%%%%%%%%%%%%%%%%%%%%%%%%%%%%%%%%%%%%%%%%%%%%%%%%%%%%%%%%%%%%%%%%%%%%%%%%%%%%%%%%%%%%%%%%%%%%%%%%%%%%%%%%%%%%%%%%%%%%%%%%%%%%
% Figure 10: Dynamic contact angles
%%%%%%%%%%%%%%%%%%%%%%%%%%%%%%%%%%%%%%%%%%%%%%%%%%%%%%%%%%%%%%%%%%%%%%%%%%%%%%%%%%%%%%%%%%%%%%%%%%%%%%%%%%%%%%%%%%%%%%%%%%%%%%%%
%%%%%%%%%%%%%%%%%%%%%%%%%%%%%%%%%%%%%%%%%%%%%%%%%%%%%%%%%%%%%%%%%%%%%%%%%%%%%%%%%%%%%%%%%%%%%%%%%%%%%%%%%%%%%%%%%%%%%%%%%%%%%%%%
%
%%%%%%%%%%%%%%%%%%%%%%%%%%%%%%%%%%%%%%%%%%%%%%%%%%%%%%%%%%%%%%%%%%%%%%
% SEC: Flow structure
%%%%%%%%%%%%%%%%%%%%%%%%%%%%%%%%%%%%%%%%%%%%%%%%%%%%%%%%%%%%%%%%%%%%%%
%
\subsection{Flow structure near a contact line}
\begin{figure}
\center
\subfigure[]{\includegraphics[angle=0,scale=0.39]{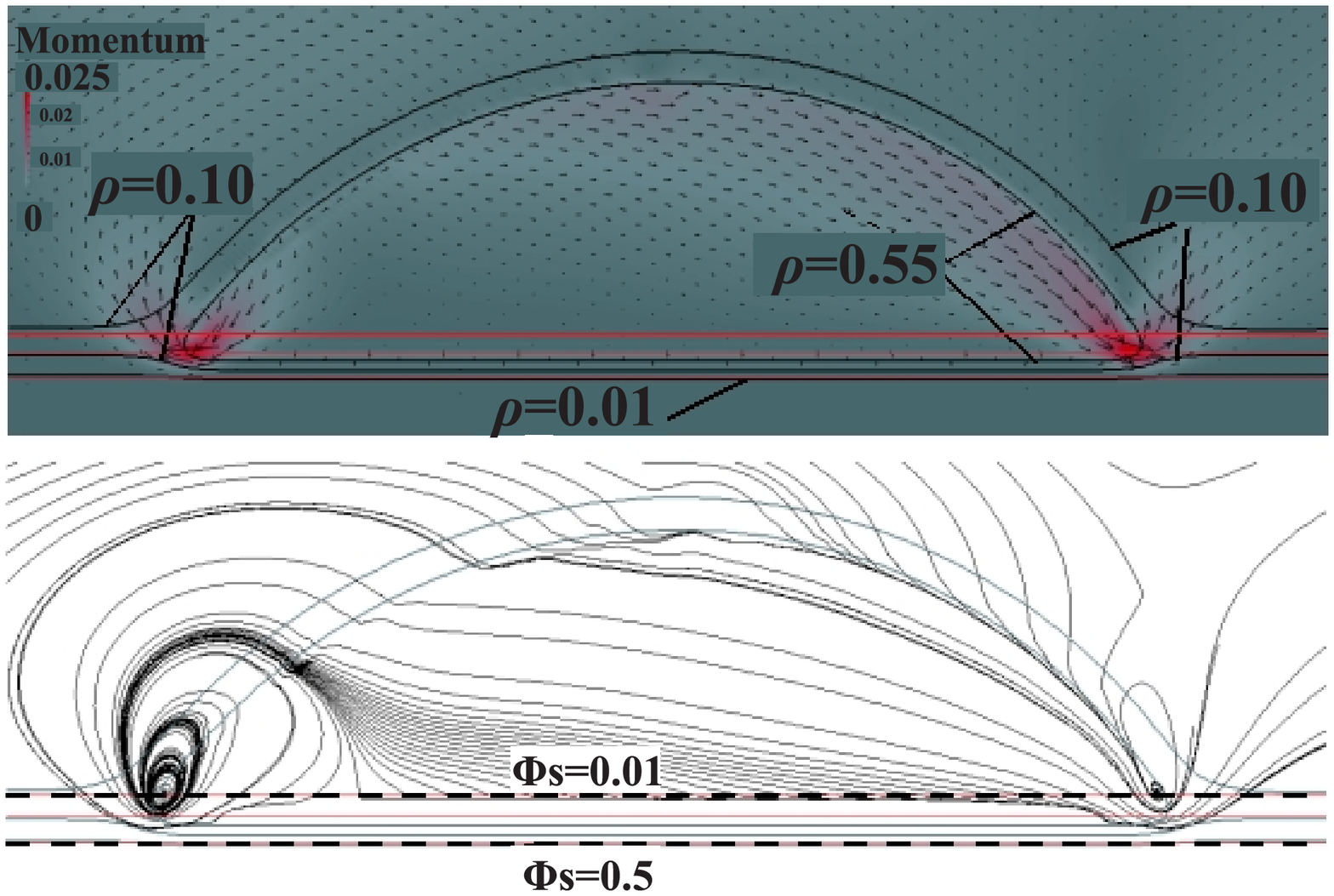}}\\
\subfigure[]{\includegraphics[angle=0,scale=0.39]{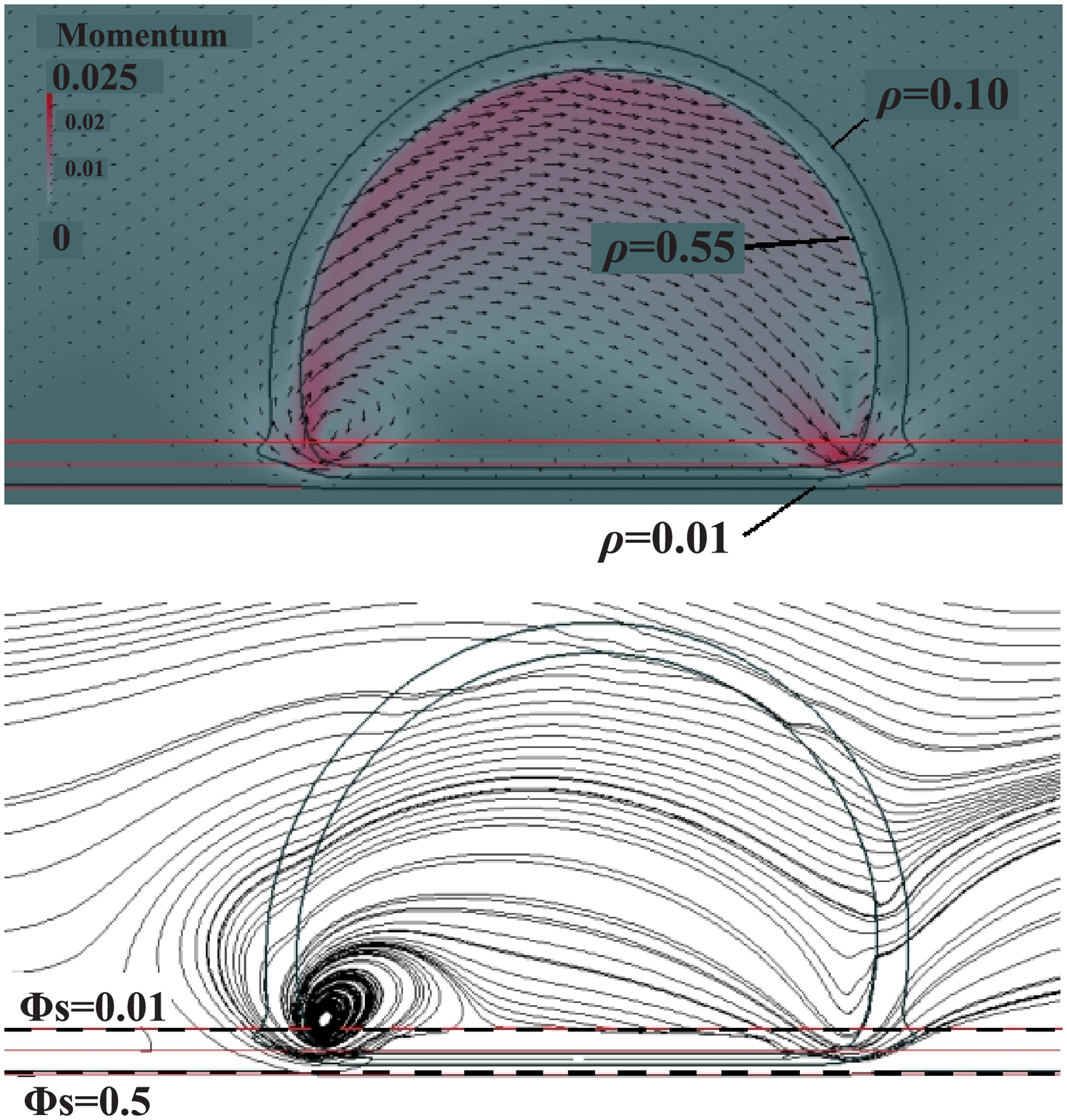}}
\caption{Fluid momentum magnitude $|{\bf M}^{(1)}|$ (top panels in the subfigures) and velocity streamlines (bottom panel in the subfigures) for droplet motion under pressure-driven flow.
(a) Static contact angle $\theta_Y=58^\circ$ ($G_R=1.33$, $G_A=0.5$, $\epsilon=1.0$). 
(b) Static contact angle $\theta_Y=113^\circ$ ($G_R=1.33$, $G_A=0.2$, $\epsilon=1.0$)
}
\label{fig:flow1}
\end{figure}

\begin{figure}
\center
\subfigure[]{\includegraphics[angle=0,scale=0.33]{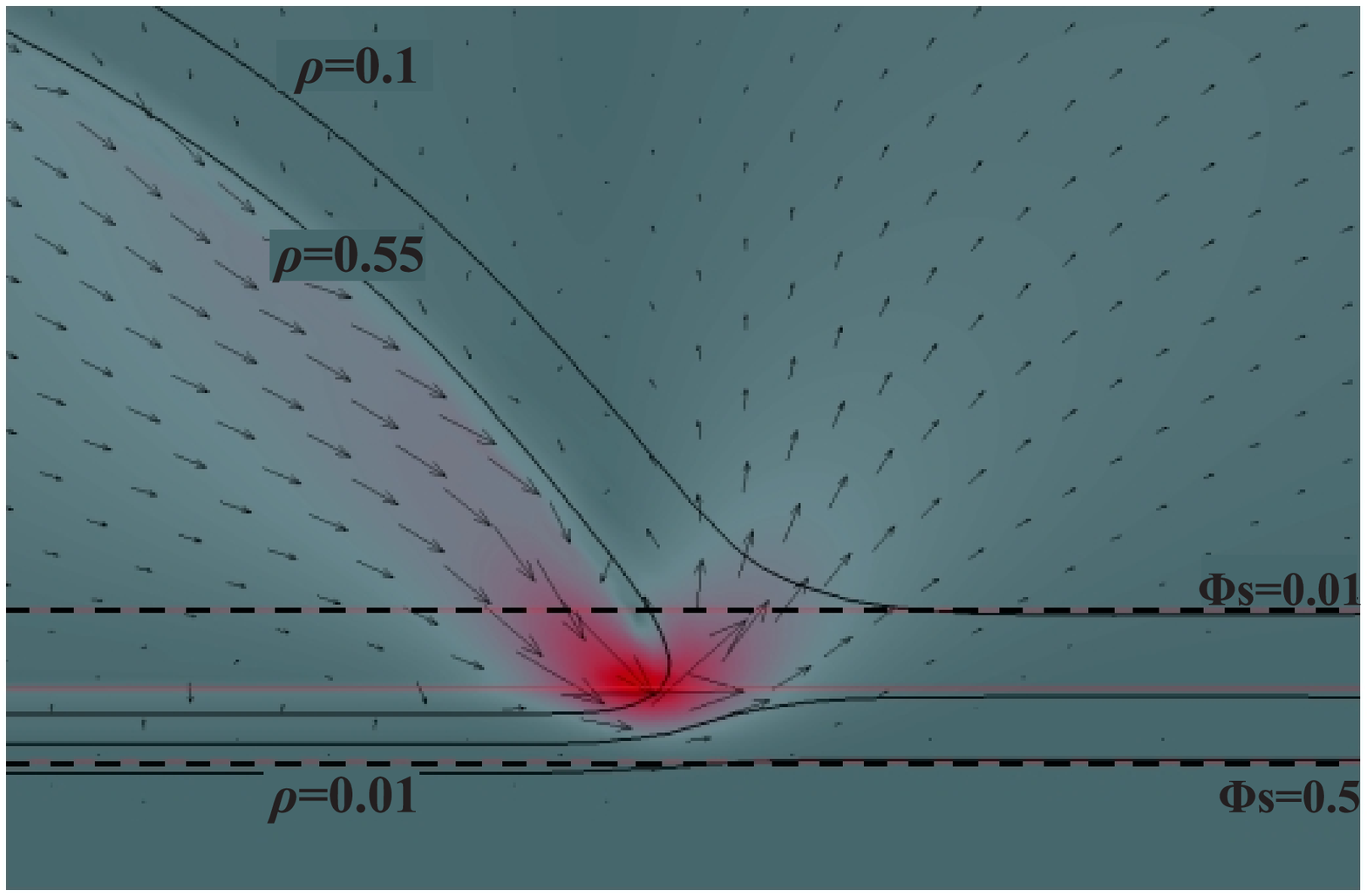}}
\subfigure[]{\includegraphics[angle=0,scale=0.33]{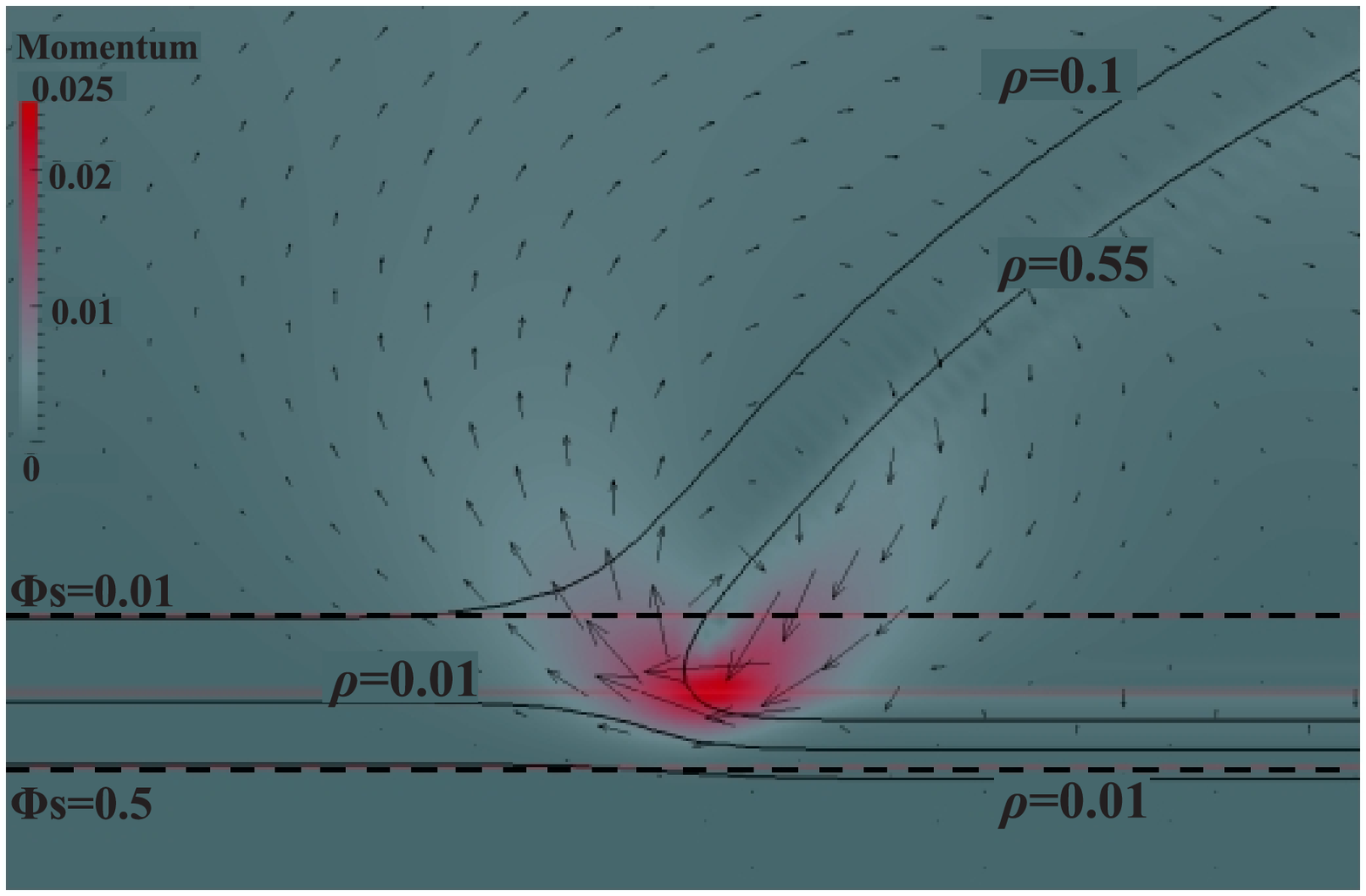}}
\subfigure[]{\includegraphics[angle=0,scale=0.33]{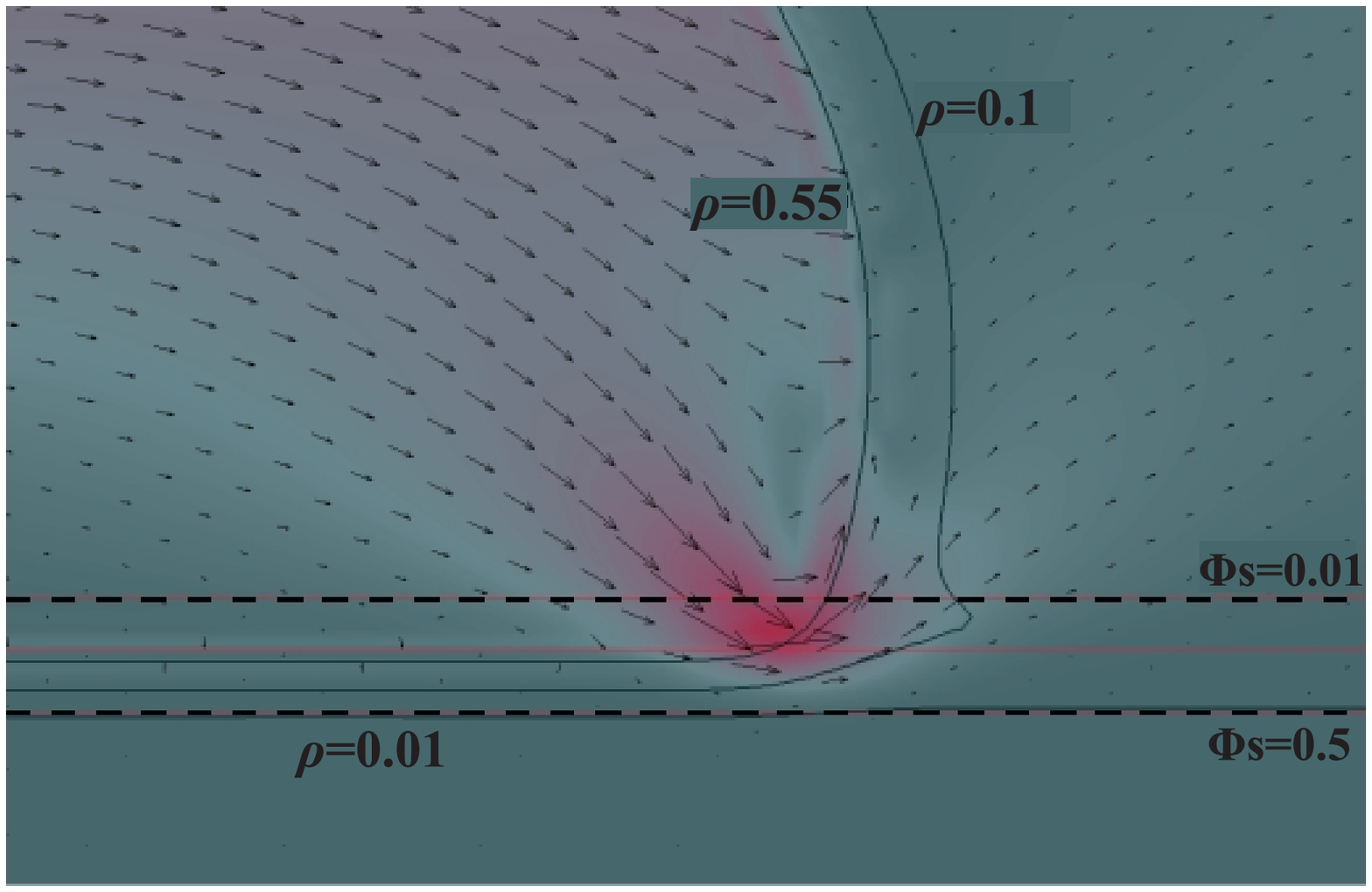}}
\subfigure[]{\includegraphics[angle=0,scale=0.33]{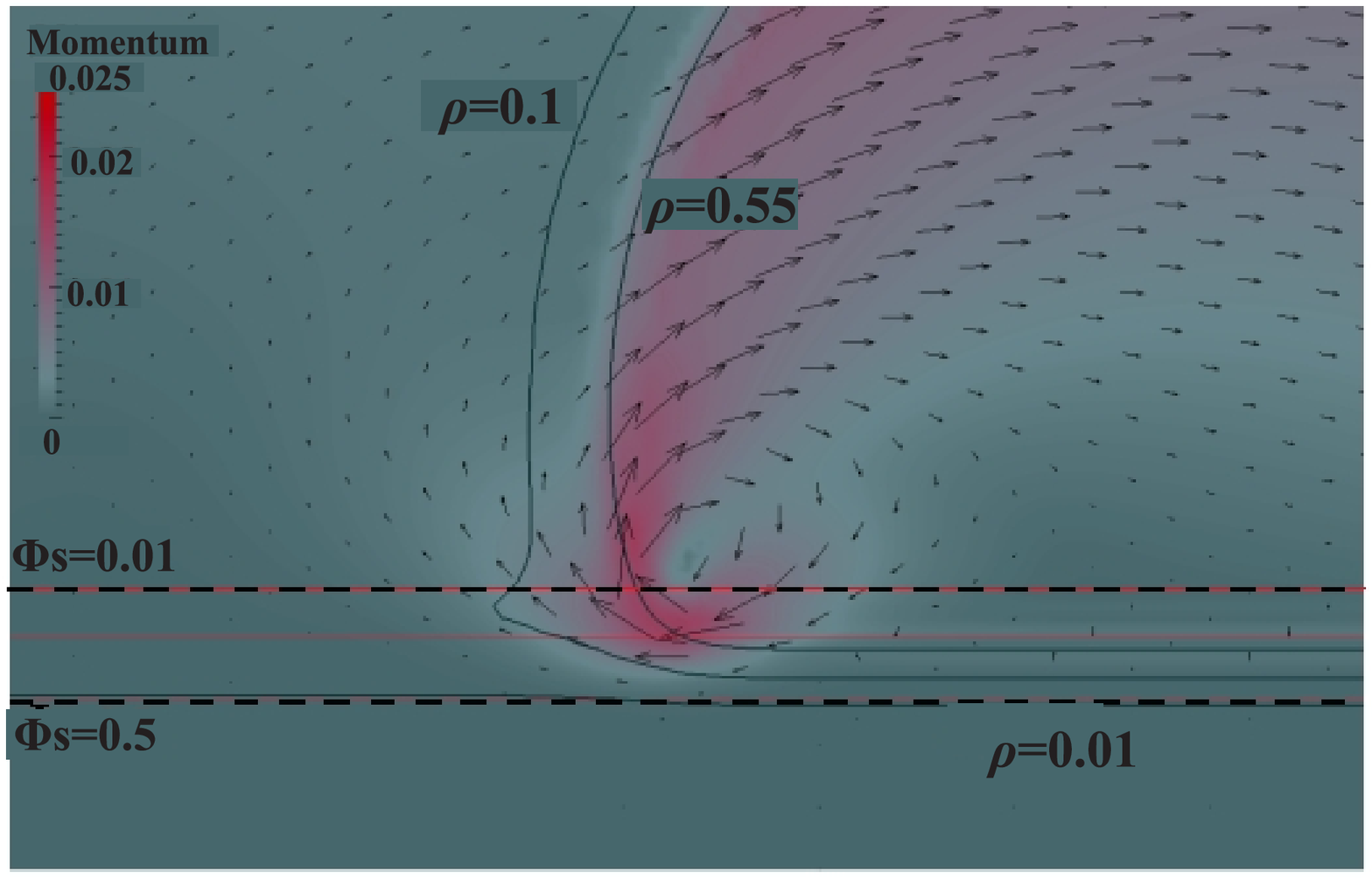}}
\caption{Flow patterns in the vicinity of the advancing and receding contact lines observed on a hydrophilic surface and on a hydrophobic surface.
(a) Advancing contact line, static contact angle $\theta_Y=58^\circ$ ($G_R=1.33$, $G_A=0.5$, $\epsilon=1.0$). (b) Receding contact line, static contact angle $\theta_Y=58^\circ$ ($G_R=1.33$, $G_A=0.5$, $\epsilon=1.0$). (c)  Advancing contact line, static contact angle $\theta_Y=113^\circ$ ($G_R=1.33$, $G_A=0.2$, $\epsilon=1.0$). (d) Receding contact line, static contact angle $\theta_Y=113^\circ$ ($G_R=1.33$, $G_A=0.2$, $\epsilon=1.0$).}
\label{fig:flow2}
\end{figure}

The flow field in the vicinity of the advancing and receding contact lines is reported in Figs.~\ref{fig:flow1}--\ref{fig:flow2}.
There are several key features produced by the present model whose existence and crucial exhibited effects have been discussed in previous works \cite{Blake1999,Lukyanov2007,Shikhmurzaev2011}.
At the apparent three-phase contact line the flow becomes more intense, while the hydrodynamic slip significantly increases.
Within the droplet bulk, away from the contact lines, little or no slip is observed at the hydrodynamic boundary $y_w$ (where $\phi_S(y_w)=0.01$).
The flow kinematics approaching the advancing contact line resembles the corner flow configuration assumed by classical Voinov-Cox models \cite{Cox1998}.

The flow at the receding contact line, however, develops a trailing vortex [see Fig.~\ref{fig:flow2}(b-c)] as the droplet moves forward.
Similar trailing vortices have been experimentally reported using PIV visualization in \cite{Sakai2009}.
The reported flow features elucidate the widely different dynamics of the advancing and receding contact angles reported in Fig.~\ref{fig:Ca_qAqR} for the modeled volatile fluid on macroscopically smooth surfaces.
The substantial difference in the flow kinematics at the receding and advancing contact lines can be thought to result from the combined effects of a large gradient of the disjoining pressure and hydrodynamics in the droplet bulk.  

We close this section with a few additional remarks.
It is widely accepted that some form of hydrodynamic slip must occur in order to allow the motion of a contact line \cite{Dussan1979,deGennes1985,Shikhmurzaev1994}.
Previous works \cite{Dussan1979,deGennes1985,Shikhmurzaev2011} pointed out, however, that simply replacing no-slip by a slip boundary condition does not preserve the experimentally observed kinematics of the flow.
Physical and conceptual issues remain in the modeling of hydrodynamic boundary conditions for moving contact lines [see \cite{Shikhmurzaev2011} for a discussion]. 
Furthermore, there is some experimental \cite{Blake1999} and theoretical \cite{Lukyanov2007} evidence that dynamic contact angles cannot be solely a function of the contact line speed and the physical properties of the system.
A mesoscopic model does not require prescribing hydrodynamic boundary conditions nor the value of dynamic contact angles and, thus constitutes an interesting alternative to study dynamic wetting phenomena. 
Furthermore, conditions within the solid phase are dynamically computed with the proposed mesoscopic model and one does not need to prescribe any variables at the fluid-solid interface (e.g. pseudo-potential value, gradient of an order parameter), as it is customary for other LB models \cite{Sbragaglia2007, Pooley2009}.
%
%
%%%%%%%%%%%%%%%%%%%%%%%%%%%%%%%%%%%%%%%%%%%%%%%%%%%%%%%%%%%%%%%%%%%%%%%%%%%%%%%%%%%%%%%%%%%%%%%%%%%%%%%%%%%%%%%%%%%%%%%%%%%%%%%%
% Figure 11: Flow field
%%%%%%%%%%%%%%%%%%%%%%%%%%%%%%%%%%%%%%%%%%%%%%%%%%%%%%%%%%%%%%%%%%%%%%%%%%%%%%%%%%%%%%%%%%%%%%%%%%%%%%%%%%%%%%%%%%%%%%%%%%%%%%%%
%
%%%%%%%%%%%%%%%%%%%%%%%%%%%%%%%%%%%%%%%%%%%%%%%%%%%%%%%%%%%%%%%%%%%%%%%%%%%%%%%%%%%%%%%%%%%%%%%%%%%%%%%%%%%%%%%%%%%%%%%%%%%%%%%%

%%%%%%%%%%%%%%%%%%%%%%%%%%%%%%%%%%%%%%%%%%%%%%%%%%%%%%%%%%%%%%%%%%%%%%%%%%%%%%%%%%%%%%%%%%%%%%%%%%%%%%%%%%%%%%%%%%%%%%%%%%%%%%%%
% Figure 12: Contact lines
%%%%%%%%%%%%%%%%%%%%%%%%%%%%%%%%%%%%%%%%%%%%%%%%%%%%%%%%%%%%%%%%%%%%%%%%%%%%%%%%%%%%%%%%%%%%%%%%%%%%%%%%%%%%%%%%%%%%%%%%%%%%%%%%
%%%%%%%%%%%%%%%%%%%%%%%%%%%%%%%%%%%%%%%%%%%%%%%%%%%%%%%%%%%%%%%%%%%%%%%%%%%%%%%%%%%%%%%%%%%%%%%%%%%%%%%%%%%%%%%%%%%%%%%%%%%%%%%%
%
%
%
\section{Summary}
We have presented a mesoscopic model, based on a statistical description of the microscopic physics, that is capable of simulating nontrivial macroscopic phenomena observed during dynamic wetting of solids when hydrodynamic effects play a critical role.
Similarly to previous models based on the lattice Boltzmann method \cite{Shan1994,Benzi2009}, long-range microscopic interactions are modeled via mean-field potentials (i.e. pseudo-potentials) that are explicit functions of macroscopic variables. 
The most distinct feature of this model is that the solid phase is treated on a similar footing as the fluid phase. 
Short-range fluid-solid interactions are modeled by a solid probability function $\phi_S$, defining the probability of fluid-solid collisions, while
long-range fluid-solid interactions are determined by pseudo-potentials designed to produce a disjoining pressure with the proper functional form.  
Hence, the modeled surface forces have essential features required to mimic physical effects observed during both full and partial wetting of solids.    
Regarding these fundamental effects we remark the following. 
(1) The effective slip velocity, determined by momentum fluxes across the surface film, can widely vary on surfaces that exhibit the same equilibrium contact angle (i.e. surfaces with similar chemical properties). 
(2) The magnitude of the hydrodynamic slip can significantly vary across macroscopic scales; the slip velocity increases near a moving contact line, while no slip is approximately recovered far from it. 
(3) Precursor films develop on hydrophilic surfaces and their evolution is dictated by the EOS of the fluid and surface forces; for the simulated system a small fraction of vapor is adsorbed on the solid substrate underneath the film.  
(4) A finite hysteresis of the static contact angle is observed for partially wettable and macroscopically smooth surfaces; the magnitude of this hysteresis is determined by the modeled shape of the disjoining pressure.
(5) The flow kinematics near advancing and receding contact lines can exhibit substantial differences; rolling motion is favored near the receding contact line, while corner flow easily develops near the advancing contact line.
The ability of the model we presented to reproduce these physical phenomena is crucial for studying diverse problems involving partial wetting and hydrodynamic interactions.
\section{Discussion and Conclusions}

There are several limitations inherent in the LB implementation that originate with the discretization of phase space.
As a consequence, the method is stable and robust within a range of moderate density and viscosity ratios between the vapor and liquid phases, while it is difficult to realize large surface tensions (i.e. above one in lattice units).
The study of many systems of interest can be accomplished by carefully matching the most relevant dimensionless groups in the problem of interest (e.g. Capillary number, Bond number, Ohnesorge number).
However, modeling of systems with large density ratios (e.g. water-air) in general flow conditions remains a challenging task and requires further developments of LB and other mesoscopic methods.
An interesting possibility is to implement a mesoscopic approach at the same level of modeling by employing a particle-based method; this could potentially extend the range of physical parameters where the numerical method is stable at the expense of certain computational advantages in the lattice discretization.

We should remark that the proposed approach aims to reproduce the proper macroscopic physics in the fluid bulk.
The purpose of our mesoscopic description is to dynamically couple surface forces and hydrodynamics, circumventing the use of complex boundary conditions for macroscopic variables.
The interfacial film thickness in the proposed model is arbitrarily prescribed by placing a minimal number of simulation nodes (i.e. one lattice cell) required to resolve the interfacial region [see Appendix].
The most attractive property of our approach lies in the fact that the chemical physics and hydrodynamics, fully coupled in the mesoscopic model, arise as different aspects of the simulated dynamics of a single-particle distribution.
This level of description has the potential of guiding the study of complex phenomena in contact line motion.
Such phenomena could include the finite time for formation and relaxation of the solid-fluid interface, and fluid compressibility effects. 
The application of this model to study deformable solids is of particular interest; there, $\phi_S$ becomes time dependent.
Other interesting applications include the dynamic wetting of micro-structured surfaces, and the extension of the proposed model to multi-component systems (e.g. colloidal suspensions, complex fluids).
\begin{acknowledgments}
The authors are pleased to acknowledge M. Sbragaglia, X. Shan, S. Succi, and S. Sundaresan  for valuable comments and stimulating discussions throughout the progress of this work.
The work has been partially supported by the U.S Department of Energy through Grant DE-FG02-09ER25877
(CEC and IGK) and by the European Research Council under the European Community's Seventh Framework Programme
(FP7 / 2007-2013) / ERC Grant Agreement no [240710] (MEK and AGP)
\end{acknowledgments}
%
%
%******************************************************************************
%***     Appendix 
%******************************************************************************
\appendix
\section{\label{app:} The wall functions}
All fluid-solid interactions are determined by the spatially-varying functions ($\phi_S$, $\bar{\psi}_R$, and $\bar{\psi}_A$) shown in Fig.~\ref{fig:solidwall}, which must be defined before the dynamic simulation starts.
In principle, the three functions can be any integrable function with a finite value in the solid bulk and null value within the fluid bulk.
The spatial variation of these functions determines the magnitude of the surface forces (i.e. disjoining pressure) and momentum fluxes in the dynamic simulation.
The wall function $\phi_S({\bf x})$ introduced in Eq.~\ref{eq:wall_shift} controls the momentum exchange at the fluid-solid interface and thus the effective slip velocity dynamically attained above the interfacial film.
A finite value of the local wall probability ($0<\phi_S<1$) can be interpreted as the presence of microscopic-scale imperfections (or sub-nanometer scale roughness) on the solid surface; 
the product $\rho \phi_S({\bf x})$ represents the local probability of a collisional event (short-range repulsion) between solid and fluid molecules.
The repulsive component, $\bar{\psi}_R$, and attractive component ,$\bar{\psi}_A$, of the Fluid-Solid pseudo-potential, $\psi_{FS}$, give rise to a disjoining pressure with both repulsive and attractive parts; this property is crucial to model partially wettable surfaces \cite{Starov2009}.

The conditions described above could be easily satisfied by simple analytical expressions in the case of flat surfaces, as we studied in this work.
However, it is convenient to have a general numerical procedure to deal with complex surface geometries.
For that purpose we employ a recursive Gaussian filter
\begin{equation}
G_{S}^{(N_s)}({\bf x})= \sum_{n_s=1}^{N_s} \sum_{i=1}^{Q} w_i \phi_{S}^{(n_s-1)}({\bf x}+{\bf r}_i).
\label{eq:iteration}
\end{equation}
The initial wall probability $G_{S}^{0}=H({\bf x})$ is a Heaviside step function that takes the unit value, $H({\bf x} \in {\Omega})=1$, in the region $\Omega$  containing the solid nodes and takes the zero value outside this region, 
$H({\bf x} \notin {\Omega})=0$.
In our simulations the outer boundary of the domain $\partial \Omega$ belongs to the solid bulk, in this region the value of the wall probability must always be unit $G_{S}^{(n_s)} |_{\partial \Omega}=1$.
Accordingly, the smoothed wall function $G_{S}^{(n_s)}$ is reset to unity on the outer boundary nodes $\partial \Omega$ after each iteration in Eq.~\ref{eq:iteration}.
All numerical results in this work employ $\bar{\psi}_{R}=G_{S}^{(5)}$, $\bar{\psi}_{A}=G_{S}^{(22)}-G_{S}^{(5)}$, $\phi_{S}=G_{S}^{(22)}$ [see Fig.~\ref{fig:solidwall}].
\newpage 
%\bibliography{lbm1}

\begin{thebibliography}{27}%
\makeatletter
\providecommand \@ifxundefined [1]{%
 \@ifx{#1\undefined}
}%
\providecommand \@ifnum [1]{%
 \ifnum #1\expandafter \@firstoftwo
 \else \expandafter \@secondoftwo
 \fi
}%
\providecommand \@ifx [1]{%
 \ifx #1\expandafter \@firstoftwo
 \else \expandafter \@secondoftwo
 \fi
}%
\providecommand \natexlab [1]{#1}%
\providecommand \enquote  [1]{``#1''}%
\providecommand \bibnamefont  [1]{#1}%
\providecommand \bibfnamefont [1]{#1}%
\providecommand \citenamefont [1]{#1}%
\providecommand \href@noop [0]{\@secondoftwo}%
\providecommand \href [0]{\begingroup \@sanitize@url \@href}%
\providecommand \@href[1]{\@@startlink{#1}\@@href}%
\providecommand \@@href[1]{\endgroup#1\@@endlink}%
\providecommand \@sanitize@url [0]{\catcode `\\12\catcode `\$12\catcode
  `\&12\catcode `\#12\catcode `\^12\catcode `\_12\catcode `\%12\relax}%
\providecommand \@@startlink[1]{}%
\providecommand \@@endlink[0]{}%
\providecommand \url  [0]{\begingroup\@sanitize@url \@url }%
\providecommand \@url [1]{\endgroup\@href {#1}{\urlprefix }}%
\providecommand \urlprefix  [0]{URL }%
\providecommand \Eprint [0]{\href }%
\providecommand \doibase [0]{http://dx.doi.org/}%
\providecommand \selectlanguage [0]{\@gobble}%
\providecommand \bibinfo  [0]{\@secondoftwo}%
\providecommand \bibfield  [0]{\@secondoftwo}%
\providecommand \translation [1]{[#1]}%
\providecommand \BibitemOpen [0]{}%
\providecommand \bibitemStop [0]{}%
\providecommand \bibitemNoStop [0]{.\EOS\space}%
\providecommand \EOS [0]{\spacefactor3000\relax}%
\providecommand \BibitemShut  [1]{\csname bibitem#1\endcsname}%
\let\auto@bib@innerbib\@empty
%</preamble>
\bibitem [{\citenamefont {Lauga}\ \emph {et~al.}(2006)\citenamefont {Lauga},
  \citenamefont {Brenner},\ and\ \citenamefont {Stone}}]{Lauga2006}%
  \BibitemOpen
  \bibfield  {author} {\bibinfo {author} {\bibfnamefont {E.}~\bibnamefont
  {Lauga}}, \bibinfo {author} {\bibfnamefont {M.}~\bibnamefont {Brenner}}, \
  and\ \bibinfo {author} {\bibfnamefont {H.}~\bibnamefont {Stone}},\
  }\href@noop {} {\bibfield  {journal} {\bibinfo  {journal} {Perspective}\
  }\textbf {\bibinfo {volume} {17}},\ \bibinfo {pages} {1} (\bibinfo {year}
  {2006})}\BibitemShut {NoStop}%
\bibitem [{\citenamefont {Shikhmurzaev}(2011)}]{Shikhmurzaev2011}%
  \BibitemOpen
  \bibfield  {author} {\bibinfo {author} {\bibfnamefont {Y.}~\bibnamefont
  {Shikhmurzaev}},\ }\href@noop {} {\bibfield  {journal} {\bibinfo  {journal}
  {Europ. Phys. Jour.}\ }\textbf {\bibinfo {volume} {197}},\ \bibinfo {pages}
  {47} (\bibinfo {year} {2011})}\BibitemShut {NoStop}%
\bibitem [{\citenamefont {Cox}(1998)}]{Cox1998}%
  \BibitemOpen
  \bibfield  {author} {\bibinfo {author} {\bibfnamefont {R.}~\bibnamefont
  {Cox}},\ }\href@noop {} {\bibfield  {journal} {\bibinfo  {journal} {J. Fluid
  Mech.}\ }\textbf {\bibinfo {volume} {357}},\ \bibinfo {pages} {249} (\bibinfo
  {year} {1998})}\BibitemShut {NoStop}%
\bibitem [{\citenamefont {Shikhmurzaev}(1994)}]{Shikhmurzaev1994}%
  \BibitemOpen
  \bibfield  {author} {\bibinfo {author} {\bibfnamefont {Y.~D.}\ \bibnamefont
  {Shikhmurzaev}},\ }\href@noop {} {\bibfield  {journal} {\bibinfo  {journal}
  {Fluid Dyn. Res.}\ }\textbf {\bibinfo {volume} {13}},\ \bibinfo {pages} {45 }
  (\bibinfo {year} {1994})}\BibitemShut {NoStop}%
\bibitem [{\citenamefont {Dussan}(1979)}]{Dussan1979}%
  \BibitemOpen
  \bibfield  {author} {\bibinfo {author} {\bibfnamefont {E.}~\bibnamefont
  {Dussan}},\ }\href@noop {} {\bibfield  {journal} {\bibinfo  {journal} {Annu.
  Rev. Fluid Mech.}\ }\textbf {\bibinfo {volume} {11}},\ \bibinfo {pages} {371}
  (\bibinfo {year} {1979})}\BibitemShut {NoStop}%
\bibitem [{\citenamefont {de~Gennes}(1985)}]{deGennes1985}%
  \BibitemOpen
  \bibfield  {author} {\bibinfo {author} {\bibfnamefont {P.~G.}\ \bibnamefont
  {de~Gennes}},\ }\href {\doibase 10.1103/RevModPhys.57.827} {\bibfield
  {journal} {\bibinfo  {journal} {Rev. Mod. Phys.}\ }\textbf {\bibinfo {volume}
  {57}},\ \bibinfo {pages} {827} (\bibinfo {year} {1985})}\BibitemShut
  {NoStop}%
\bibitem [{\citenamefont {Kralchevsky}\ and\ \citenamefont
  {Nagayama}(2001)}]{Book1}%
  \BibitemOpen
  \bibfield  {author} {\bibinfo {author} {\bibfnamefont {P.}~\bibnamefont
  {Kralchevsky}}\ and\ \bibinfo {author} {\bibfnamefont {K.}~\bibnamefont
  {Nagayama}},\ }\href@noop {} {\emph {\bibinfo {title} {Particles at fluids
  interfaces and membranes}}},\ Vol.~\bibinfo {volume} {10}\ (\bibinfo
  {publisher} {Elsevier Science},\ \bibinfo {year} {2001})\BibitemShut
  {NoStop}%
\bibitem [{\citenamefont {Cieplak}\ \emph {et~al.}(2001)\citenamefont
  {Cieplak}, \citenamefont {Koplik},\ and\ \citenamefont
  {Banavar}}]{Koplik2001}%
  \BibitemOpen
  \bibfield  {author} {\bibinfo {author} {\bibfnamefont {M.}~\bibnamefont
  {Cieplak}}, \bibinfo {author} {\bibfnamefont {J.}~\bibnamefont {Koplik}}, \
  and\ \bibinfo {author} {\bibfnamefont {J.~R.}\ \bibnamefont {Banavar}},\
  }\href@noop {} {\bibfield  {journal} {\bibinfo  {journal} {Phys. Rev. Lett.}\
  }\textbf {\bibinfo {volume} {86}},\ \bibinfo {pages} {803} (\bibinfo {year}
  {2001})}\BibitemShut {NoStop}%
\bibitem [{\citenamefont {Starov}\ and\ \citenamefont
  {Velarde}(2009)}]{Starov2009}%
  \BibitemOpen
  \bibfield  {author} {\bibinfo {author} {\bibfnamefont {V.}~\bibnamefont
  {Starov}}\ and\ \bibinfo {author} {\bibfnamefont {M.}~\bibnamefont
  {Velarde}},\ }\href@noop {} {\bibfield  {journal} {\bibinfo  {journal} {J.
  Phys.: Cond. Matt.}\ }\textbf {\bibinfo {volume} {21}},\ \bibinfo {pages}
  {464121} (\bibinfo {year} {2009})}\BibitemShut {NoStop}%
\bibitem [{\citenamefont {Shan}\ and\ \citenamefont {Chen}(1994)}]{Shan1994}%
  \BibitemOpen
  \bibfield  {author} {\bibinfo {author} {\bibfnamefont {X.}~\bibnamefont
  {Shan}}\ and\ \bibinfo {author} {\bibfnamefont {H.}~\bibnamefont {Chen}},\
  }\href@noop {} {\bibfield  {journal} {\bibinfo  {journal} {Phys. Rev. E}\
  }\textbf {\bibinfo {volume} {49}},\ \bibinfo {pages} {2941} (\bibinfo {year}
  {1994})}\BibitemShut {NoStop}%
\bibitem [{\citenamefont {Shan}\ \emph {et~al.}(2006)\citenamefont {Shan},
  \citenamefont {Yuan},\ and\ \citenamefont {Chen}}]{Shan2006}%
  \BibitemOpen
  \bibfield  {author} {\bibinfo {author} {\bibfnamefont {X.}~\bibnamefont
  {Shan}}, \bibinfo {author} {\bibfnamefont {X.}~\bibnamefont {Yuan}}, \ and\
  \bibinfo {author} {\bibfnamefont {H.}~\bibnamefont {Chen}},\ }\href@noop {}
  {\bibfield  {journal} {\bibinfo  {journal} {J. Fluid Mech.}\ }\textbf
  {\bibinfo {volume} {550}},\ \bibinfo {pages} {413} (\bibinfo {year}
  {2006})}\BibitemShut {NoStop}%
\bibitem [{\citenamefont {Benzi}\ \emph {et~al.}(2009)\citenamefont {Benzi},
  \citenamefont {Chibbaro},\ and\ \citenamefont {Succi}}]{Benzi2009}%
  \BibitemOpen
  \bibfield  {author} {\bibinfo {author} {\bibfnamefont {R.}~\bibnamefont
  {Benzi}}, \bibinfo {author} {\bibfnamefont {S.}~\bibnamefont {Chibbaro}}, \
  and\ \bibinfo {author} {\bibfnamefont {S.}~\bibnamefont {Succi}},\
  }\href@noop {} {\bibfield  {journal} {\bibinfo  {journal} {Phys. Rev. Lett.}\
  }\textbf {\bibinfo {volume} {102}},\ \bibinfo {pages} {026002} (\bibinfo
  {year} {2009})}\BibitemShut {NoStop}%
\bibitem [{\citenamefont {Deryagin}(1955)}]{Deryagin1955}%
  \BibitemOpen
  \bibfield  {author} {\bibinfo {author} {\bibfnamefont {B.}~\bibnamefont
  {Deryagin}},\ }\href@noop {} {\bibfield  {journal} {\bibinfo  {journal}
  {Colloid J. USSR}\ }\textbf {\bibinfo {volume} {17}},\ \bibinfo {pages} {191}
  (\bibinfo {year} {1955})}\BibitemShut {NoStop}%
\bibitem [{\citenamefont {Teletzke}\ \emph {et~al.}(1988)\citenamefont
  {Teletzke}, \citenamefont {Davis}, \citenamefont {Scriven} \emph
  {et~al.}}]{Teletzke1988}%
  \BibitemOpen
  \bibfield  {author} {\bibinfo {author} {\bibfnamefont {G.}~\bibnamefont
  {Teletzke}}, \bibinfo {author} {\bibfnamefont {H.}~\bibnamefont {Davis}},
  \bibinfo {author} {\bibfnamefont {L.}~\bibnamefont {Scriven}},  \emph
  {et~al.},\ }\href@noop {} {\bibfield  {journal} {\bibinfo  {journal} {Rev.
  Phys. Appl.}\ }\textbf {\bibinfo {volume} {23}},\ \bibinfo {pages} {989}
  (\bibinfo {year} {1988})}\BibitemShut {NoStop}%
\bibitem [{\citenamefont {Colosqui}(2010)}]{colosqui2010}%
  \BibitemOpen
  \bibfield  {author} {\bibinfo {author} {\bibfnamefont {C.~E.}~\bibnamefont
  {Colosqui}},\ }\href@noop {} {\bibfield  {journal} {\bibinfo  {journal}
  {Phys. Rev. E}\ }\textbf {\bibinfo {volume} {81}},\ \bibinfo {pages} {026702}
  (\bibinfo {year} {2010})}\BibitemShut {NoStop}%
\bibitem [{\citenamefont {Shan}(2008)}]{Shan2008}%
  \BibitemOpen
  \bibfield  {author} {\bibinfo {author} {\bibfnamefont {X.}~\bibnamefont
  {Shan}},\ }\href@noop {} {\bibfield  {journal} {\bibinfo  {journal} {Phys.
  Rev. E}\ }\textbf {\bibinfo {volume} {77}},\ \bibinfo {pages} {066702}
  (\bibinfo {year} {2008})}\BibitemShut {NoStop}%
\bibitem [{Note1()}]{Note1}%
  \BibitemOpen
  \bibinfo {note} {Simulations in this work required 1 to 4 hours of CPU time
  in a mid-size worksation (quad-core processor at 2.8Ghz and 8GB of
  RAM)}\BibitemShut {NoStop}%
\bibitem [{\citenamefont {Zhao}(2008)}]{ZhaoGPU}%
  \BibitemOpen
  \bibfield  {author} {\bibinfo {author} {\bibfnamefont {Y.}~\bibnamefont
  {Zhao}},\ }\href@noop {} {\bibfield  {journal} {\bibinfo  {journal} {The
  Visual Computer}\ }\textbf {\bibinfo {volume} {24}},\ \bibinfo {pages} {323}
  (\bibinfo {year} {2008})}\BibitemShut {NoStop}%
\bibitem [{\citenamefont {Bernaschi}\ \emph {et~al.}(2010)\citenamefont
  {Bernaschi}, \citenamefont {Fatica}, \citenamefont {Melchionna},
  \citenamefont {Succi},\ and\ \citenamefont {Kaxiras}}]{BernaschiGPU}%
  \BibitemOpen
  \bibfield  {author} {\bibinfo {author} {\bibfnamefont {M.}~\bibnamefont
  {Bernaschi}}, \bibinfo {author} {\bibfnamefont {M.}~\bibnamefont {Fatica}},
  \bibinfo {author} {\bibfnamefont {S.}~\bibnamefont {Melchionna}}, \bibinfo
  {author} {\bibfnamefont {S.}~\bibnamefont {Succi}}, \ and\ \bibinfo {author}
  {\bibfnamefont {E.}~\bibnamefont {Kaxiras}},\ }\href@noop {} {\bibfield
  {journal} {\bibinfo  {journal} {Concurrency and Computation: Practice and
  Experience}\ }\textbf {\bibinfo {volume} {22}},\ \bibinfo {pages} {1}
  (\bibinfo {year} {2010})}\BibitemShut {NoStop}%
\bibitem [{\citenamefont {Kupershtokh}\ \emph {et~al.}(2009)\citenamefont
  {Kupershtokh}, \citenamefont {Medvedev},\ and\ \citenamefont
  {Karpov}}]{Kupershtokh2009}%
  \BibitemOpen
  \bibfield  {author} {\bibinfo {author} {\bibfnamefont {A.}~\bibnamefont
  {Kupershtokh}}, \bibinfo {author} {\bibfnamefont {D.}~\bibnamefont
  {Medvedev}}, \ and\ \bibinfo {author} {\bibfnamefont {D.}~\bibnamefont
  {Karpov}},\ }\href@noop {} {\bibfield  {journal} {\bibinfo  {journal} {Comp.
  Math. Appl.}\ }\textbf {\bibinfo {volume} {58}},\ \bibinfo {pages} {965}
  (\bibinfo {year} {2009})}\BibitemShut {NoStop}%
\bibitem [{\citenamefont {Colosqui}\ \emph {et~al.}(2012)\citenamefont
  {Colosqui}, \citenamefont {Falcucci}, \citenamefont {Ubertini},\ and\
  \citenamefont {Succi}}]{colosqui2012}%
  \BibitemOpen
  \bibfield  {author} {\bibinfo {author} {\bibfnamefont {C.~E.}~\bibnamefont
  {Colosqui}}, \bibinfo {author} {\bibfnamefont {G.}~\bibnamefont {Falcucci}},
  \bibinfo {author} {\bibfnamefont {S.}~\bibnamefont {Ubertini}}, \ and\
  \bibinfo {author} {\bibfnamefont {S.}~\bibnamefont {Succi}},\ }\href@noop {}
  {\bibfield  {journal} {\bibinfo  {journal} {Soft matter}\ }\textbf {\bibinfo
  {volume} {In press}} (\bibinfo {year} {2012})}\BibitemShut {NoStop}%
\bibitem [{\citenamefont {Sbragaglia}\ \emph {et~al.}(2007)\citenamefont
  {Sbragaglia}, \citenamefont {Benzi}, \citenamefont {Biferale}, \citenamefont
  {Succi}, \citenamefont {Sugiyama},\ and\ \citenamefont
  {Toschi}}]{Sbragaglia2007}%
  \BibitemOpen
  \bibfield  {author} {\bibinfo {author} {\bibfnamefont {M.}~\bibnamefont
  {Sbragaglia}}, \bibinfo {author} {\bibfnamefont {R.}~\bibnamefont {Benzi}},
  \bibinfo {author} {\bibfnamefont {L.}~\bibnamefont {Biferale}}, \bibinfo
  {author} {\bibfnamefont {S.}~\bibnamefont {Succi}}, \bibinfo {author}
  {\bibfnamefont {K.}~\bibnamefont {Sugiyama}}, \ and\ \bibinfo {author}
  {\bibfnamefont {F.}~\bibnamefont {Toschi}},\ }\href@noop {} {\bibfield
  {journal} {\bibinfo  {journal} {Phys. Rev. E}\ }\textbf {\bibinfo {volume}
  {75}},\ \bibinfo {pages} {026702} (\bibinfo {year} {2007})}\BibitemShut
  {NoStop}%
\bibitem [{\citenamefont {Pooley}\ \emph {et~al.}(2009)\citenamefont {Pooley},
  \citenamefont {Kusumaatmaja},\ and\ \citenamefont {Yeomans}}]{Pooley2009}%
  \BibitemOpen
  \bibfield  {author} {\bibinfo {author} {\bibfnamefont {C.~M.}\ \bibnamefont
  {Pooley}}, \bibinfo {author} {\bibfnamefont {H.}~\bibnamefont
  {Kusumaatmaja}}, \ and\ \bibinfo {author} {\bibfnamefont {J.~M.}\
  \bibnamefont {Yeomans}},\ }\href@noop {} {\bibfield  {journal} {\bibinfo
  {journal} {Eur. Phys. J. - Spec. Top.}\ }\textbf {\bibinfo {volume} {171}},\
  \bibinfo {pages} {63} (\bibinfo {year} {2009})}\BibitemShut {NoStop}%
\bibitem [{\citenamefont {Mognetti}\ \emph {et~al.}(2010)\citenamefont
  {Mognetti}, \citenamefont {Kusumaatmaja},\ and\ \citenamefont
  {Yeomans}}]{Mognetti2010}%
  \BibitemOpen
  \bibfield  {author} {\bibinfo {author} {\bibfnamefont {B.~M.}\ \bibnamefont
  {Mognetti}}, \bibinfo {author} {\bibfnamefont {H.}~\bibnamefont
  {Kusumaatmaja}}, \ and\ \bibinfo {author} {\bibfnamefont {J.~M.}\
  \bibnamefont {Yeomans}},\ }\href@noop {} {\bibfield  {journal} {\bibinfo
  {journal} {Faraday Discuss.}\ }\textbf {\bibinfo {volume} {146}},\ \bibinfo
  {pages} {153} (\bibinfo {year} {2010})}\BibitemShut {NoStop}%
\bibitem [{\citenamefont {Blake}\ \emph {et~al.}(1999)\citenamefont {Blake},
  \citenamefont {Bracke},\ and\ \citenamefont {Shikhmurzaev}}]{Blake1999}%
  \BibitemOpen
  \bibfield  {author} {\bibinfo {author} {\bibfnamefont {T.}~\bibnamefont
  {Blake}}, \bibinfo {author} {\bibfnamefont {M.}~\bibnamefont {Bracke}}, \
  and\ \bibinfo {author} {\bibfnamefont {Y.}~\bibnamefont {Shikhmurzaev}},\
  }\href@noop {} {\bibfield  {journal} {\bibinfo  {journal} {Phys. Fluids}\
  }\textbf {\bibinfo {volume} {11}},\ \bibinfo {pages} {1995} (\bibinfo {year}
  {1999})}\BibitemShut {NoStop}%
\bibitem [{\citenamefont {Lukyanov}\ and\ \citenamefont
  {Shikhmurzaev}(2007)}]{Lukyanov2007}%
  \BibitemOpen
  \bibfield  {author} {\bibinfo {author} {\bibfnamefont {A.~V.}\ \bibnamefont
  {Lukyanov}}\ and\ \bibinfo {author} {\bibfnamefont {Y.~D.}\ \bibnamefont
  {Shikhmurzaev}},\ }\href@noop {} {\bibfield  {journal} {\bibinfo  {journal}
  {Phys. Rev. E}\ }\textbf {\bibinfo {volume} {75}},\ \bibinfo {pages} {051604}
  (\bibinfo {year} {2007})}\BibitemShut {NoStop}%
\bibitem [{\citenamefont {Sakai}\ \emph {et~al.}(2009)\citenamefont {Sakai},
  \citenamefont {Kono}, \citenamefont {Nakajima}, \citenamefont {Sakai},
  \citenamefont {Abe},\ and\ \citenamefont {Fujishima}}]{Sakai2009}%
  \BibitemOpen
  \bibfield  {author} {\bibinfo {author} {\bibfnamefont {M.}~\bibnamefont
  {Sakai}}, \bibinfo {author} {\bibfnamefont {H.}~\bibnamefont {Kono}},
  \bibinfo {author} {\bibfnamefont {A.}~\bibnamefont {Nakajima}}, \bibinfo
  {author} {\bibfnamefont {H.}~\bibnamefont {Sakai}}, \bibinfo {author}
  {\bibfnamefont {M.}~\bibnamefont {Abe}}, \ and\ \bibinfo {author}
  {\bibfnamefont {A.}~\bibnamefont {Fujishima}},\ }\href@noop {} {\bibfield
  {journal} {\bibinfo  {journal} {Lang.}\ }\textbf {\bibinfo {volume} {26}},\
  \bibinfo {pages} {1493} (\bibinfo {year} {2009})}\BibitemShut {NoStop}%
\end{thebibliography}
\end{document}